\documentclass[aps,prd,preprintnumbers,superscriptaddress,nofootinbib,amssymb,notitlepage,eqsecnum]{revtex4-1}

\usepackage{here}
\usepackage{graphicx}
\usepackage{amsmath}
\usepackage{placeins}
\usepackage{amssymb}
\usepackage{amsfonts}
\usepackage{bm}
\usepackage{color}
\usepackage[dvipsnames]{xcolor}
\usepackage{fvextra}  
\usepackage{framed}
\usepackage{fontspec}
\usepackage{multirow}
\makeatletter
\renewcommand{\p@subsection}{}
\makeatother
\newcounter{subsubsubsection}[subsubsection]

\usepackage{comment}
\usepackage{tikz}
\usetikzlibrary{decorations.pathreplacing,arrows.meta}
\usepackage{enumitem}
\usepackage[colorlinks=true,hyperfootnotes=true,citecolor=cyan,linkcolor=blue,urlcolor=blue]{hyperref}

\newcommand{\Hy}{\ensuremath{\mathrm{H}}} 
\newcommand{\He}{\ensuremath{\mathrm{He}}} 
\newcommand{\ini}{\ensuremath{\mathrm{ini}}} 
\newcommand\bwd[1]{\overset{\leftarrow}{#1}}  
\newcommand\fwd[1]{\overset{\rightarrow}{#1}} 

\usepackage{xspace}

\newcommand{\SymBoltz}{\textnormal{SymBoltz}\xspace}
\newcommand{\CAMB}{\textnormal{CAMB}\xspace}
\newcommand{\CLASS}{\textnormal{CLASS}\xspace}

\definecolor{refs}{RGB}{245,156,74}
\definecolor{codebackground}{RGB}{248,248,248}
\definecolor{codeborder}{RGB}{190,190,190}
\colorlet{shadecolor}{codebackground}
\makeatletter

\makeatletter
\def\PY@reset{\let\PY@it=\relax \let\PY@bf=\relax%
    \let\PY@ul=\relax \let\PY@tc=\relax%
    \let\PY@bc=\relax \let\PY@ff=\relax}
\def\PY@tok#1{\csname PY@tok@#1\endcsname}
\def\PY@toks#1+{\ifx\relax#1\empty\else%
    \PY@tok{#1}\expandafter\PY@toks\fi}
\def\PY@do#1{\PY@bc{\PY@tc{\PY@ul{%
    \PY@it{\PY@bf{\PY@ff{#1}}}}}}}
\def\PY#1#2{\PY@reset\PY@toks#1+\relax+\PY@do{#2}}

\@namedef{PY@tok@w}{\def\PY@tc##1{\textcolor[rgb]{0.73,0.73,0.73}{##1}}}
\@namedef{PY@tok@c}{\let\PY@it=\textit\def\PY@tc##1{\textcolor[rgb]{0.24,0.48,0.48}{##1}}}
\@namedef{PY@tok@cp}{\def\PY@tc##1{\textcolor[rgb]{0.61,0.40,0.00}{##1}}}
\@namedef{PY@tok@k}{\let\PY@bf=\textbf\def\PY@tc##1{\textcolor[rgb]{0.00,0.50,0.00}{##1}}}
\@namedef{PY@tok@kp}{\def\PY@tc##1{\textcolor[rgb]{0.00,0.50,0.00}{##1}}}
\@namedef{PY@tok@kt}{\def\PY@tc##1{\textcolor[rgb]{0.69,0.00,0.25}{##1}}}
\@namedef{PY@tok@o}{\def\PY@tc##1{\textcolor[rgb]{0.40,0.40,0.40}{##1}}}
\@namedef{PY@tok@ow}{\let\PY@bf=\textbf\def\PY@tc##1{\textcolor[rgb]{0.67,0.13,1.00}{##1}}}
\@namedef{PY@tok@nb}{\def\PY@tc##1{\textcolor[rgb]{0.00,0.50,0.00}{##1}}}
\@namedef{PY@tok@nf}{\def\PY@tc##1{\textcolor[rgb]{0.00,0.00,1.00}{##1}}}
\@namedef{PY@tok@nc}{\let\PY@bf=\textbf\def\PY@tc##1{\textcolor[rgb]{0.00,0.00,1.00}{##1}}}
\@namedef{PY@tok@nn}{\let\PY@bf=\textbf\def\PY@tc##1{\textcolor[rgb]{0.00,0.00,1.00}{##1}}}
\@namedef{PY@tok@ne}{\let\PY@bf=\textbf\def\PY@tc##1{\textcolor[rgb]{0.80,0.25,0.22}{##1}}}
\@namedef{PY@tok@nv}{\def\PY@tc##1{\textcolor[rgb]{0.10,0.09,0.49}{##1}}}
\@namedef{PY@tok@no}{\def\PY@tc##1{\textcolor[rgb]{0.53,0.00,0.00}{##1}}}
\@namedef{PY@tok@nl}{\def\PY@tc##1{\textcolor[rgb]{0.46,0.46,0.00}{##1}}}
\@namedef{PY@tok@ni}{\let\PY@bf=\textbf\def\PY@tc##1{\textcolor[rgb]{0.44,0.44,0.44}{##1}}}
\@namedef{PY@tok@na}{\def\PY@tc##1{\textcolor[rgb]{0.41,0.47,0.13}{##1}}}
\@namedef{PY@tok@nt}{\let\PY@bf=\textbf\def\PY@tc##1{\textcolor[rgb]{0.00,0.50,0.00}{##1}}}
\@namedef{PY@tok@nd}{\def\PY@tc##1{\textcolor[rgb]{0.67,0.13,1.00}{##1}}}
\@namedef{PY@tok@s}{\def\PY@tc##1{\textcolor[rgb]{0.73,0.13,0.13}{##1}}}
\@namedef{PY@tok@sd}{\let\PY@it=\textit\def\PY@tc##1{\textcolor[rgb]{0.73,0.13,0.13}{##1}}}
\@namedef{PY@tok@si}{\let\PY@bf=\textbf\def\PY@tc##1{\textcolor[rgb]{0.64,0.35,0.47}{##1}}}
\@namedef{PY@tok@se}{\let\PY@bf=\textbf\def\PY@tc##1{\textcolor[rgb]{0.67,0.36,0.12}{##1}}}
\@namedef{PY@tok@sr}{\def\PY@tc##1{\textcolor[rgb]{0.64,0.35,0.47}{##1}}}
\@namedef{PY@tok@ss}{\def\PY@tc##1{\textcolor[rgb]{0.10,0.09,0.49}{##1}}}
\@namedef{PY@tok@sx}{\def\PY@tc##1{\textcolor[rgb]{0.00,0.50,0.00}{##1}}}
\@namedef{PY@tok@m}{\def\PY@tc##1{\textcolor[rgb]{0.40,0.40,0.40}{##1}}}
\@namedef{PY@tok@gh}{\let\PY@bf=\textbf\def\PY@tc##1{\textcolor[rgb]{0.00,0.00,0.50}{##1}}}
\@namedef{PY@tok@gu}{\let\PY@bf=\textbf\def\PY@tc##1{\textcolor[rgb]{0.50,0.00,0.50}{##1}}}
\@namedef{PY@tok@gd}{\def\PY@tc##1{\textcolor[rgb]{0.63,0.00,0.00}{##1}}}
\@namedef{PY@tok@gi}{\def\PY@tc##1{\textcolor[rgb]{0.00,0.52,0.00}{##1}}}
\@namedef{PY@tok@gr}{\def\PY@tc##1{\textcolor[rgb]{0.89,0.00,0.00}{##1}}}
\@namedef{PY@tok@ge}{\let\PY@it=\textit}
\@namedef{PY@tok@gs}{\let\PY@bf=\textbf}
\@namedef{PY@tok@ges}{\let\PY@bf=\textbf\let\PY@it=\textit}
\@namedef{PY@tok@gp}{\let\PY@bf=\textbf\def\PY@tc##1{\textcolor[rgb]{0.00,0.00,0.50}{##1}}}
\@namedef{PY@tok@go}{\def\PY@tc##1{\textcolor[rgb]{0.44,0.44,0.44}{##1}}}
\@namedef{PY@tok@gt}{\def\PY@tc##1{\textcolor[rgb]{0.00,0.27,0.87}{##1}}}
\@namedef{PY@tok@err}{\def\PY@bc##1{{\setlength{\fboxsep}{\string -\fboxrule}\fcolorbox[rgb]{1.00,0.00,0.00}{1,1,1}{\strut ##1}}}}
\@namedef{PY@tok@kc}{\let\PY@bf=\textbf\def\PY@tc##1{\textcolor[rgb]{0.00,0.50,0.00}{##1}}}
\@namedef{PY@tok@kd}{\let\PY@bf=\textbf\def\PY@tc##1{\textcolor[rgb]{0.00,0.50,0.00}{##1}}}
\@namedef{PY@tok@kn}{\let\PY@bf=\textbf\def\PY@tc##1{\textcolor[rgb]{0.00,0.50,0.00}{##1}}}
\@namedef{PY@tok@kr}{\let\PY@bf=\textbf\def\PY@tc##1{\textcolor[rgb]{0.00,0.50,0.00}{##1}}}
\@namedef{PY@tok@bp}{\def\PY@tc##1{\textcolor[rgb]{0.00,0.50,0.00}{##1}}}
\@namedef{PY@tok@fm}{\def\PY@tc##1{\textcolor[rgb]{0.00,0.00,1.00}{##1}}}
\@namedef{PY@tok@vc}{\def\PY@tc##1{\textcolor[rgb]{0.10,0.09,0.49}{##1}}}
\@namedef{PY@tok@vg}{\def\PY@tc##1{\textcolor[rgb]{0.10,0.09,0.49}{##1}}}
\@namedef{PY@tok@vi}{\def\PY@tc##1{\textcolor[rgb]{0.10,0.09,0.49}{##1}}}
\@namedef{PY@tok@vm}{\def\PY@tc##1{\textcolor[rgb]{0.10,0.09,0.49}{##1}}}
\@namedef{PY@tok@sa}{\def\PY@tc##1{\textcolor[rgb]{0.73,0.13,0.13}{##1}}}
\@namedef{PY@tok@sb}{\def\PY@tc##1{\textcolor[rgb]{0.73,0.13,0.13}{##1}}}
\@namedef{PY@tok@sc}{\def\PY@tc##1{\textcolor[rgb]{0.73,0.13,0.13}{##1}}}
\@namedef{PY@tok@dl}{\def\PY@tc##1{\textcolor[rgb]{0.73,0.13,0.13}{##1}}}
\@namedef{PY@tok@s2}{\def\PY@tc##1{\textcolor[rgb]{0.73,0.13,0.13}{##1}}}
\@namedef{PY@tok@sh}{\def\PY@tc##1{\textcolor[rgb]{0.73,0.13,0.13}{##1}}}
\@namedef{PY@tok@s1}{\def\PY@tc##1{\textcolor[rgb]{0.73,0.13,0.13}{##1}}}
\@namedef{PY@tok@mb}{\def\PY@tc##1{\textcolor[rgb]{0.40,0.40,0.40}{##1}}}
\@namedef{PY@tok@mf}{\def\PY@tc##1{\textcolor[rgb]{0.40,0.40,0.40}{##1}}}
\@namedef{PY@tok@mh}{\def\PY@tc##1{\textcolor[rgb]{0.40,0.40,0.40}{##1}}}
\@namedef{PY@tok@mi}{\def\PY@tc##1{\textcolor[rgb]{0.40,0.40,0.40}{##1}}}
\@namedef{PY@tok@il}{\def\PY@tc##1{\textcolor[rgb]{0.40,0.40,0.40}{##1}}}
\@namedef{PY@tok@mo}{\def\PY@tc##1{\textcolor[rgb]{0.40,0.40,0.40}{##1}}}
\@namedef{PY@tok@ch}{\let\PY@it=\textit\def\PY@tc##1{\textcolor[rgb]{0.24,0.48,0.48}{##1}}}
\@namedef{PY@tok@cm}{\let\PY@it=\textit\def\PY@tc##1{\textcolor[rgb]{0.24,0.48,0.48}{##1}}}
\@namedef{PY@tok@cpf}{\let\PY@it=\textit\def\PY@tc##1{\textcolor[rgb]{0.24,0.48,0.48}{##1}}}
\@namedef{PY@tok@c1}{\let\PY@it=\textit\def\PY@tc##1{\textcolor[rgb]{0.24,0.48,0.48}{##1}}}
\@namedef{PY@tok@cs}{\let\PY@it=\textit\def\PY@tc##1{\textcolor[rgb]{0.24,0.48,0.48}{##1}}}

\def\PYZus{\char`\_}

\def\PYZca{\char`\^}

\def\PYZsh{\char`\#}

\def\PYZhy{\char`\-}

\makeatother

\makeatother

\newcommand{\LCDM}{$\Lambda$CDM}
\newcommand{\dd}{{\rm d}}

\newcommand{\mH}{{\mathcal H}}
\newcommand{\Qb}{\bar Q}
\newcommand{\rhob}{\bar\rho}
\newcommand{\Pb}{\bar P}

\newcommand{\be}{\begin{equation}}
\newcommand{\ee}{\end{equation}}
\newcommand{\bea}{\begin{eqnarray}}
\newcommand{\eea}{\end{eqnarray}}
\newcommand{\bse}{\begin{subequations}}
\newcommand{\ese}{\end{subequations}}

\usepackage{array}
\newcommand{\rowtag}{\refstepcounter{equation}\text{(\theequation)}}

\begin{document}

\title{Simple multi-interacting fluids in SymBoltz}
\author{David Figueruelo}
\email{david.figueruelo@ehu.eus}
\affiliation{Department of Theoretical Physics, University of the Basque Country UPV/EHU, 48080 Bilbao, Spain}
\affiliation{Institute of Theoretical Astrophysics, University of Oslo, N-0315 Oslo, Norway}

\author{Herman Sletmoen}
\email{herman.sletmoen@astro.uio.no}
\affiliation{Institute of Theoretical Astrophysics, University of Oslo, N-0315 Oslo, Norway}
\date{\today}

\begin{abstract}
We implement a simple and general way of specifying arbitrary interactions between baryons, dark matter and dark energy in the Einstein--Boltzmann code \SymBoltz.
Due to the symbolic-numeric interface and approximation-free structure of \SymBoltz,
the model definition is remarkably compact and convenient to modify by specifying only the covariant coupling.
The full equations are stiff, but stable and fast to integrate with an efficient implicit solver,
particularly with the sparse and analytical ODE Jacobian generated from the symbolic equations.
We first validate it using single-interaction models, comparing their background evolution and matter and CMB power spectra with those from modified versions of \CLASS, finding $0.1\%$--$1\%$ agreement at most scales.
Finally, we show how our framework makes it trivial to implement  models with any interactions of any complexity  among all three species simultaneously, which requires a more involved implementation in other Einstein--Boltzmann codes like \CLASS or \CAMB.
Our full model is specified in only around 300 lines of code and is easy to modify for any other interaction by just specifying it in three lines.
\end{abstract}

\maketitle

\vspace{-0.75cm}
\section{Introduction}

\SymBoltz\footnote{\url{https://github.com/hersle/SymBoltz.jl/}} \cite{Sletmoen:2025fro} is a recently introduced Einstein--Boltzmann solver written in the Julia programming language.
It is completely free of approximation schemes that complicate other codes such as \CAMB~\cite{2000ApJ...538..473L,Lewis:2026mif} and \CLASS~\cite{2011arXiv1104.2932L,2011JCAP...07..034B},
and it has a symbolic-numeric interface that makes it convenient to specify and modify equations.
These features are useful for implementing models with interactions between multiple species.

First, approximation-freeness means that only the full and stiff Einstein--Boltzmann equations have to be specified,
and it is not necessary to derive and implement approximations to the equations in different regimes.
In \CLASS, for example, the tight-coupling approximation (TCA)~\cite{2011JCAP...07..034B} is mandatory to circumvent stiffness of the equations in the early universe.
Interacting models generally introduce additional stiffness and modify the equations underlying the TCA,
requiring new or modified approximation schemes.
Other approximations such as the radiation streaming approximation (RSA), ultra-relativistic fluid approximation (UFA) and non-cold dark matter fluid approximation (NCDMFA) \cite{2011JCAP...07..034B,2011JCAP...09..032L} are in principle optional and can be disabled, but are often necessary in practice because the code relies on them for the speed of solving the perturbation equations.
\SymBoltz instead uses the efficient implicit Rodas5P ODE solver \cite{Rodas5P} to handle stiffness, and provides it with fast numerical code for the ODE and its sparse and analytical Jacobian generated by the symbolic-numeric interface.
This makes it possible to directly implement the full modified equations once, without spending effort on re-deriving, re-implementing and re-validating approximation schemes in different regimes.

Second, the symbolic-numeric interface allows for a convenient, compact and general way of specifying our modified equations.
Our full \LCDM{} model, with the option to add interactions, is encoded in only around 300 lines of code.%
\footnote{\url{https://hersle.github.io/SymBoltz.jl/stable/interacting/}}
All variables, parameters and equations for the background and perturbations and for all components are specified in a single place using Unicode characters with one-to-one correspondence with the mathematical notation.
The symbolic-numeric engine automatically separates the background and perturbations, splines the background when solving the perturbations, generates optimal numerical code for the ODE and its Jacobian and makes it trivial to obtain any variables from the solution.
Moreover, \SymBoltz has been updated to version 2 alongside this paper,
which makes it possible to integrate the background equations using a mixture of forwards and backwards integration in time.
This is particularly useful for interacting models,
where the background continuity equations generally lose their analytical solutions, and it is useful to integrate them backwards from present values of the density parameters. Thus, once an interaction has been defined covariantly, its implementation in \SymBoltz reduces to three symbolic lines of code. This same prescription applies to any arbitrary interaction, to any number of coupled species and/or simultaneous interactions, making even complicated interacting cosmologies a compact modification.

This paper is organised as follows. Section~\ref{sec:SB_v2} describes the new background integration framework introduced in \SymBoltz version 2. Section~\ref{sec:theory} establishes the covariant formulation of  interactions used. Section~\ref{sec:implementation} explains their implementation. Section~\ref{sec:validation} presents the validation against \CLASS. Section~\ref{sec:multiple_interactions} presents a realistic application as a prototype use case. Finally, Section~\ref{sec:conclusions} summarises our conclusions.

\begin{figure}
\centering
\begin{tikzpicture}[
    var/.style = {draw, thick, circle, minimum size=0.9cm, inner sep=1pt},
    s1/.style = {var, draw=blue!70!black, fill=blue!12},
    s2/.style = {var, draw=violet!70!black, fill=violet!12},
    s3/.style = {var, draw=orange!80!black, fill=orange!15},
    s4/.style = {var, draw=red!70!black, fill=red!12},
    s5/.style = {var, draw=green!50!black, fill=green!12},
    dep/.style = {-{Stealth}, gray!70},
    cpl/.style = {{Stealth}-{Stealth}, gray!70},
    leg/.style = {minimum size=0.35cm, inner sep=0pt},
]
    \def\R{2.6} 

    \node[s1] (rhob)  at ( 90:\R) {$\bwd{\rho_{\rm b}}$};
    \node[s1] (rhoc)  at ( 50:\R) {$\bwd{\rho_{\rm c}}$};
    \node[s1] (rhoL)  at ( 10:\R) {$\bwd{\rho_\Lambda}$};
    \node[s2] (chi)   at (330:\R) {$\bwd{\chi}$};
    \node[s3] (tau)   at (290:\R) {$\fwd{\tau}$};
    \node[s4] (xH)    at (250:\R) {$\fwd{x_\Hy^+}$};
    \node[s4] (xHe)   at (210:\R) {$\fwd{x_\He^+}$};
    \node[s4] (Tb)    at (170:\R) {$\fwd{T_{\rm b}}$};
    \node[s5] (kappa) at (130:\R) {$\bwd{\kappa}$};

    \foreach \s in {rhob, rhoc, rhoL}
        \foreach \t in {chi, tau, xH, xHe, Tb, kappa}
            \draw[dep] (\s) -- (\t);
    \foreach \s in {xH, xHe, Tb}
        \draw[dep] (\s) -- (kappa);

    \draw[cpl] (rhob) -- (rhoc);
    \draw[cpl] (rhoc) -- (rhoL);
    \draw[cpl] (rhob) -- (rhoL);
    \draw[cpl] (xH) -- (xHe);
    \draw[cpl] (xHe) -- (Tb);
    \draw[cpl] (xH) -- (Tb);

    \foreach \n/\a in {rhob/90, rhoc/50, rhoL/10, xH/250, xHe/210, Tb/170}
        \draw[dep] (\n.center) ++(\a:0.55) ++(\a+232:0.22)
            arc[start angle=\a+232, end angle=\a+488, radius=0.22];

    \node[s1, leg] (l1) at (4.0,  1.2) {};
    \node[right] at (l1.east) {Stage 1 ($t_\ini \leftarrow t_0$ backwards)};
    \node[s2, leg] (l2) at (4.0,  0.6) {};
    \node[right] at (l2.east) {Stage 2 ($t_\ini \leftarrow t_0$ backwards)};
    \node[s3, leg] (l3) at (4.0,  0.0) {};
    \node[right] at (l3.east) {Stage 3 ($t_\ini \rightarrow t_0$ forwards)};
    \node[s4, leg] (l4) at (4.0, -0.6) {};
    \node[right] at (l4.east) {Stage 4 ($t_\ini \rightarrow t_0$ forwards)};
    \node[s5, leg] (l5) at (4.0, -1.2) {};
    \node[right] at (l5.east) {Stage 5 ($t_\ini \leftarrow t_0$ backwards)};
\end{tikzpicture}
\caption{\SymBoltz version 2 automatically creates a dependency graph between all variables in the user-specified background equations.
An arrow $x \rightarrow y$ means that $y^\prime$ (and hence $y$) depends on $x$.
\SymBoltz identifies cycles in this graph (here with the same color),
which must be integrated as one coupled ODE.
Different cycles are disconnected or connected by one-way arrows,
which constitute stages of the equations that can be integrated sequentially,
each using splined input of earlier stages.
Variables without outgoing arrows correspond to integrals.
Each variable is marked with a preferred integration direction (forwards $\rightarrow$ or backwards $\leftarrow$),
and all variables in the same coupled stage must have the same direction.
Here the background system \eqref{eq:bg} admits 5 stages (going clockwise from 12 o'clock),
and stages $1+2$ and $3+4$ can be merged because they share direction.}
\label{fig:graph}
\end{figure}
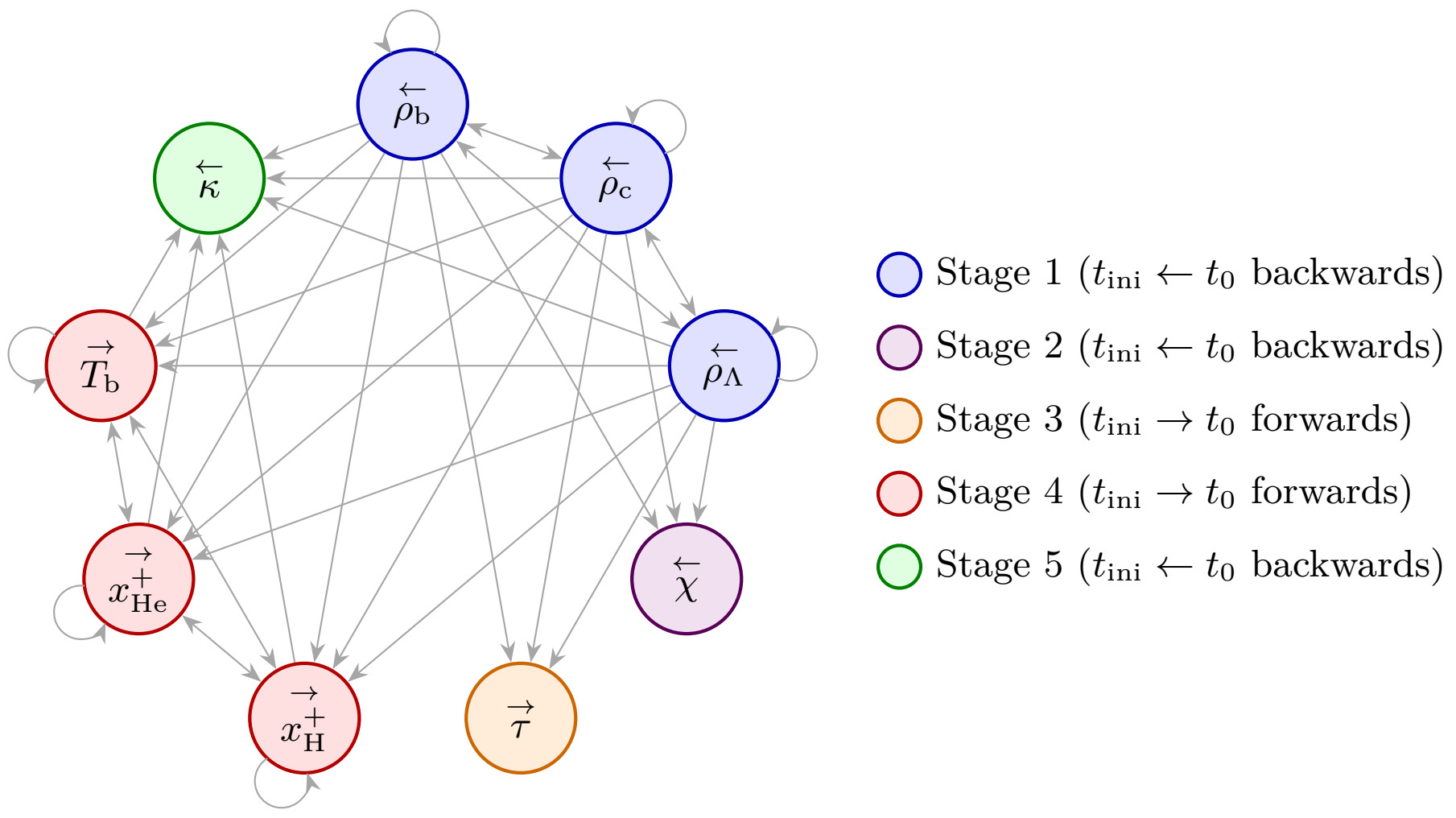

\section{SymBoltz version 2}
\label{sec:SB_v2}

\SymBoltz version 2 is released alongside this paper.
In particular, it updates the symbolic-numeric framework to let users easily specify models where some background\footnote{By background we mean everything to 0th order in perturbation theory. This includes what is referred to as thermodynamics in \CLASS.} equations are integrated forwards in time, while others are integrated backwards.
For example, the background ODEs (including thermodynamics and RECFAST recombination) we consider here have the following structure
(in this section ${}^\prime = \mathrm{d}/\mathrm{d}t$ and $t=\ln a$ is the time variable used in the code, but we will later formulate our equations with conformal time):%
\begin{subequations}
\label{eq:bg}
\begin{align}
    & \text{Baryon density:} & \rho_{\rm b}{}^\prime(t) &= f(t, \rho_{\rm b}, \rho_{\rm c}, \rho_\Lambda) && \text{integrated from} && \rho_{\rm b}(t_0) && \text{and backwards}, \label{eq:rhob} \\
    & \text{Dark matter density:} & \rho_{\rm c}{}^\prime(t) &= f(t, \rho_{\rm b}, \rho_{\rm c}, \rho_\Lambda) && \text{integrated from} && \rho_{\rm c}(t_0) && \text{and backwards}, \label{eq:rhoc} \\
    & \text{Dark energy density:} & \rho_\Lambda{}^\prime(t) &= f(t, \rho_{\rm b}, \rho_{\rm c}, \rho_\Lambda) && \text{integrated from} && \rho_\Lambda(t_0) && \text{and backwards}, \label{eq:rhoL} \\
    & \text{Lookback time:} & \chi{}^\prime(t) &= f(t, \rho_{\rm b}, \rho_{\rm c}, \rho_\Lambda) && \text{integrated from} && \chi(t_0) && \text{and backwards}, \label{eq:chi} \\
    & \text{Conformal time:} & \tau{}^\prime(t) &= f(t, \rho_{\rm b}, \rho_{\rm c}, \rho_\Lambda) && \text{integrated from} && \tau(t_\ini) && \text{and forwards}, \label{eq:tau} \\
    & \text{Hydrogen ionization:} & x_\Hy^+{}^\prime(t) &= f(t, \rho_{\rm b}, \rho_{\rm c}, \rho_\Lambda, x_\Hy^+, x_\He^+, T_{\rm b}) && \text{integrated from} && x_\Hy^+(t_\ini) && \text{and forwards}, \label{eq:xH} \\
    & \text{Helium ionization:} & x_\He^+{}^\prime(t) &= f(t, \rho_{\rm b}, \rho_{\rm c}, \rho_\Lambda, x_\Hy^+, x_\He^+, T_{\rm b}) && \text{integrated from} && x_\He^+(t_\ini) && \text{and forwards}, \label{eq:xHe} \\
    & \text{Baryon temperature:} & T_{\rm b}{}^\prime(t) &= f(t, \rho_{\rm b}, \rho_{\rm c}, \rho_\Lambda, x_\Hy^+, x_\He^+, T_{\rm b}) && \text{integrated from} && T_{\rm b}(t_\ini) && \text{and forwards}, \label{eq:Tb} \\
    & \text{Optical depth:} & \kappa{}^\prime(t) &= f(t, \rho_{\rm b}, \rho_{\rm c}, \rho_\Lambda, x_\Hy^+, x_\He^+, T_{\rm b}) && \text{integrated from} && \kappa(t_0) && \text{and backwards}. \label{eq:kappa}
\end{align}
\end{subequations}
The continuity equations for $i \in \{b, c, \Lambda\}$%
\footnote{We use the label $\Lambda$ to denote dark energy of any form, including but not limited to a cosmological constant. This abuse of notation lets us denote dark energy with one consistent character and as closely as possible to the notation in \SymBoltz.}
are numerically integrated backwards because we parameterise them by their present density parameters $\Omega_{b0}$, $\Omega_{c0}$ and $\Omega_{\Lambda 0}$.
Conversely, thermodynamics equations are integrated forwards
because their fully ionised initial conditions are in the early universe.
This is a generalisation of the equations for the standard \LCDM{} model,
where the non-interacting continuity equations have analytical solutions $\rho_i(t) = \rho_i(t_0) a(t)^{-3(1+w_i)}$ with constant equations of state $w_i$ in the background.
As illustrated in Fig.~\ref{fig:graph},
the background system \eqref{eq:bg} admits five stages,
and the best way to integrate them is to join stages $1+2$ and $3+4$:
\begin{equation}
\text{integrate $\{\rho_{\rm b}, \rho_{\rm c}, \rho_\Lambda,\chi\}$ backwards}
\ \  \rightarrow \ \ 
\text{integrate $\{\tau, x_\He^+, x_\Hy^+, T_{\rm b}\}$ forwards}
\ \  \rightarrow \ \ 
\text{integrate $\{\kappa\}$ backwards}.
\end{equation}

With the latest \SymBoltz update,
the user only needs to write down the full background (and perturbation) equations together,
and the symbolic-numeric framework automatically inspects them and finds the dependency graph between all variables.
If the user requests some variables to be integrated backwards and others forwards,
the code will check if the inter-dependencies allow the system to be split in this way,
and if so split the background ODEs into appropriate stages.
This can be understood as a generalisation of the $\texttt{background.c} \to \texttt{thermodynamics.c} \to \texttt{...}$ structure in \CLASS \cite{2011arXiv1104.2932L},
such that the number of background stages is not hardcoded to 2.
We emphasise that this machinery is not custom-tailored to one particular model,
but made to be general with the ambition of supporting equations with arbitrary mixtures of forward-backward integration.

For example, \CLASS supports non-interacting continuity equations with $w = w(a)$ by first evaluating
$\rho(a_\ini) = \rho(1) \exp \left( -3 \int_1^{a_\ini} {\rm d} a \, \big(1+w(a)\big) / a \right)$ backwards,
and then integrating the continuity equation forwards as an ODE.
This falls naturally into the general backwards-forwards integration framework in \SymBoltz by first doing backwards integration on the integral.

The alternative to integrating backwards is to only integrate forwards,
but use the shooting method to find the initial conditions at $t_\ini$ that give the desired densities at $t_0$.
This is also supported in \SymBoltz.
It amounts to solving the following root-finding problem for $\rho_{\rm b}(t_\ini)$, $\rho_{\rm c}(t_\ini)$ and $\rho_\Lambda(t_\ini)$,
where $\rho_{\rm b}(t_0)$, $\rho_{\rm c}(t_0)$ and $\rho_\Lambda(t_0)$ are computed by integrating an ODE forwards:
\begin{equation}
\mathbf{F}\big(\rho_{\rm b}(t_\ini), \rho_{\rm c}(t_\ini), \rho_\Lambda(t_\ini)\big) =
\begin{bmatrix}
\rho_{\rm b}(t_0) / \rho_{\rm cr}(t_0) - \Omega_{b0}, &
\rho_{\rm c}(t_0) / \rho_{\rm cr}(t_0) - \Omega_{c0}, &
\rho_\Lambda(t_0) / \rho_{\rm cr}(t_0) - \Omega_{\Lambda 0}
\end{bmatrix}
= \mathbf{0}.
\label{eq:shoot}
\end{equation}
Here the differentiability of \SymBoltz is used to compute the Jacobian of $\mathbf{F}$ to help Newton methods converge to the right root.
Whether backwards integration or shooting is better is model-dependent and comes down to
where the initial conditions are known,
the stability of the integration direction of the equations
and the stability of the parameter space of the model.
For the interacting models here we find backward and forward integration of the continuity equations to be equally stable,
but with the shooting method we find it challenging to find initial guesses for $\rho_i(t_\ini)$ that avoid running into physically unstable regions in parameter space (e.g. with negative energy densities) for general interactions.
Shooting must also solve the background several times, while backwards integration solves it only once.
Our preferred approach is therefore backwards integration to avoid shooting.

The ambition with \SymBoltz version 2 is that
mixed backwards-forward integration in conjunction with the shooting method
constitutes a framework for treating the boundary-value nature of the background equations for many cosmological models in a general way.

\section{Theory}
\label{sec:theory}

We consider scalar perturbations in the Newtonian gauge, following \cite{Ma:1995ey}:
\be
    \dd s^2=a^2(\tau)\left[-(1+2\Psi)\dd\tau^2+(1-2\Phi)\delta_{ij}\dd x^i\dd x^j\right] \,.
\ee
Here $a$ is the scale factor, $\Phi$ and $\Psi$ are the Bardeen potentials, $\tau$ is conformal time and we use comoving spatial coordinates $x^i$.
Primes will denote derivatives with respect to conformal time $\tau$,
and the conformal Hubble rate is $\mH = a'/a$.
The energy-momentum tensor for a generic matter component $(n)$ is:
\bse
\bea
    T^0_{\ 0(n)} &=& -(\rhob_{({\rm n})}+\delta\rho_{({\rm n})})\,, \\
    T^0_{\ i(n)} &=& (\rhob_{({\rm n})}+\Pb_{({\rm n})})v_{i(n)}\,, \\
    T^i_{\ 0(n)} &=& -(\rhob_{({\rm n})}+\Pb_{({\rm n})})v^i_{({\rm n})}\,, \\
    T^i_{\ j(n)} &=& (\Pb_{({\rm n})}+\delta P_{({\rm n})})\delta^i_{\ j}+\Pb_{({\rm n})}\Pi^i_{\ j(n)}\, .
\eea
\ese
The background has energy density $\rhob_{({\rm n})}$ and pressure $\Pb_{({\rm n})}$,
while $\delta\rho_{(n)}$ and $\delta P_{(n)}$ are first order density and pressure perturbations.
We consider perfect fluids with vanishing anisotropic stress tensor \(\Pi^i_{\ j(n)}=0\), although the formalism can be extended to include anisotropic stresses.
The quantity $v^i_{(n)}$ is the fluid velocity perturbation,
which is commonly defined in terms of a velocity potential:
\be
    v_{i(n)}=-\partial_i v_{({\rm n})},
    \qquad
    \theta_{({\rm n})}\equiv i\vec k\cdot\vec v_{({\rm n})} \qquad \text{and} \qquad u^\mu_{({\rm n})}=\frac{1}{a}\left( 1-\Psi, v^{i}_{(\rm n)} \right)\; .
\label{eq:velocity}
\ee
For each component $(n)$, we define an equation of state $w$, density contrast $\delta$, velocity potential $\theta$, and the physical sound speed in the fluid rest frame (rf):
\be
    w_{({\rm n})}\equiv\frac{\Pb_{({\rm n})}}{\rhob_{({\rm n})}},
    \qquad
    \delta_{({\rm n})}\equiv\frac{\delta\rho_{({\rm n})}}{\rhob_{({\rm n})}},
    \qquad
    \theta_{({\rm n})}\equiv i\vec k\cdot\vec v_{({\rm n})} ,
    \qquad
     c_{{\rm s},({\rm n})}^2\equiv\left.\frac{\delta P_{({\rm n})}}{\delta\rho_{({\rm n})}}\right|_{\rm rf} \;.
\ee
In an arbitrary (non rest) frame we have the relation
\be
    \delta P_{({\rm n})}
    =
    c_{{\rm s},({\rm n})}^2\delta\rho_{({\rm n})}
    -
    (c_{{\rm s},({\rm n})}^2-c_{{\rm a},({\rm n})}^2)\rhob_{({\rm n})}' v_{({\rm n})} \;,
\ee
where the adiabatic sound speed is defined as
\be
c_{{\rm a},({\rm n})}^2
\equiv
\frac{\Pb_{({\rm n})}'}{\rhob_{({\rm n})}'}
=
w_{({\rm n})}
+
\frac{w_{({\rm n})}'}{\rhob_{({\rm n})}'/\rhob_{({\rm n})}}
\qquad
\xrightarrow{\;\text{non-interacting}\;}
\qquad
w_{({\rm n})}
-
\frac{w_{({\rm n})}'}{3\mH\left(1+w_{({\rm n})}\right)} \,.
\label{eq:soundspeed_adiabatic}
\ee

Having specified the metric, fluid variables and sound speed conventions, we can now write the evolution equations in the absence of interactions.
In this case, each component conserves its energy-momentum tensor independently with \(\nabla_\nu T^{\mu\nu}_{(n)}=0\).
In the background, this leads to the continuity equation
\be
    \rhob_{({\rm n})}'
    +
    3\mH(1+w_{({\rm n})})\rhob_{({\rm n})}
    =
    0 .
\ee
At the perturbative level, it leads to the density contrast and Euler equations for each perfect fluid component%
\footnote{For baryons, the Euler equation has an additional term from Thomson scattering with photons of the form $+\frac{4\rho_\gamma}{3\rho_{\rm (b)}}\,a n_e\sigma_{\rm T}\,(\theta_\gamma-\theta_{\rm (b)})$, which accounts for the momentum exchange between the two fluids. Moreover, for baryons $w_{\rm (b)}\simeq c_{{\rm a},({\rm b})}^2\simeq0$, while the sound speed $c_{{\rm s},({\rm b})}^2$ is small but nonzero. The standard treatment (theoretically~\cite{Ma:1995ey} and in codes~\cite{2011arXiv1104.2932L,2011JCAP...07..034B,Lewis:2026mif}) retains only the pressure gradient term $k^2c_{{\rm s},({\rm b})}^2\delta_{\rm (b)}$, whereas the terms $-3\mathcal{H}c_{{\rm s},({\rm b})}^2\delta_{\rm (b)}$, $-9\mathcal{H}^2c_{{\rm s},({\rm b})}^2\theta_{\rm (b)}/k^2$ and $+3\mathcal{H}c_{{\rm s},({\rm b})}^2\theta_{\rm (b)}$ are usually neglected, being suppressed by $c_{{\rm s},({\rm b})}^2$ with respect to the terms of order $\mathcal{H}$. In any case, the effect of neglecting them is extremely small and below any potential effect on any observable.}
~\cite{Ma:1995ey}
\bea
\delta_{({\rm n})}'
&=&
3\mH(w_{({\rm n})}-c_{{\rm s},({\rm n})}^2)\delta_{({\rm n})}
-(1+w_{({\rm n})})\theta_{({\rm n})}
+3(1+w_{({\rm n})})\Phi'
-
9\mH^2(c_{{\rm s},({\rm n})}^2-c_{{\rm a},({\rm n})}^2)(1+w_{({\rm n})})\frac{\theta_{({\rm n})}}{k^2}\,,\\
 \theta_{({\rm n})}'
    &=&
    -\mH(1-3c_{{\rm s},({\rm n})}^2)\theta_{({\rm n})}
    +
    \frac{k^2c_{{\rm s},({\rm n})}^2}{1+w_{({\rm n})}}\delta_{({\rm n})}
    +
    k^2\Psi \,.
\
\label{eq:delta_nonint}
\eea
However, these equations change when we add interactions through a covariantly defined transfer four-vector $Q^\mu_{(n)}$.

\subsection{Covariant formulation of interactions}
\label{subsec:cov_couplings}
Coupling terms introduced at the level of the background continuity equation are a useful phenomenological starting point.
However, this description is not enough once perturbations are included, since a covariant formulation is needed.
We therefore introduce the interaction for each component $(n)$ through a four-vector source:
\be
    \nabla_\nu T^{\mu\nu}_{({\rm n})}=Q^\mu_{({\rm n})}\,,
    \qquad
    \sum_n Q^\mu_{({\rm n})}=0 \,.
\label{eq:covQ}
\ee
The second condition guarantees total energy-momentum conservation.
For later convenience we decompose the four-vector source as~\cite{Gavela:2010tm,Valiviita:2008iv,vonMarttens:2018iav}
\bse
\bea
    Q^\mu_{({\rm n})} &=& Q_{({\rm n})}\hat u^\mu+F_{({\rm n})}^\mu\,, \\
    Q_{({\rm n})} &=& \Qb_{({\rm n})}+\delta Q_{({\rm n})}\,, \\
    F_{({\rm n})}^\mu &=& \frac{1}{a}(0,\vec f_{({\rm n})})
    =
    \frac{1}{a}(0,-\partial^j f_{({\rm n})})\,,
\eea%
\label{eq:Qdecomp}%
\ese
with $   \hat u_\mu F_{({\rm n})}^\mu=0$.
Here $\Qb_{({\rm n})}$ is the background energy-transfer rate, $\delta Q_{({\rm n})}$ is its perturbation, and $f_{({\rm n})}$ is the scalar momentum transfer potential.
The vector $\hat u^\mu$ defines the reference frame in which the energy transfer is measured,
and need not coincide with the four-velocity of any fluid.
Following the velocity conventions \eqref{eq:velocity}, we write it as
\be
    \hat u^\mu=\frac{1}{a}(1-\Psi,\hat v^i),
    \qquad
    \hat\theta\equiv i\vec k\cdot\hat{\vec v} .
\label{eq:fourvelocityref}
\ee
At the level of each matter component $(n)$, the above decomposition of the interaction translates into
\be
Q_{0\,(n)}
=
-a\left[(1+\Psi)\Qb_{({\rm n})}+\delta Q_{({\rm n})}\right],
\qquad
Q_{j\,(n)}
=
-a\partial_j\left(\Qb_{({\rm n})}\hat v+f_{({\rm n})}\right).
\label{eq:Qcomponents}
\ee
The condition $\sum_n Q^\mu_{({\rm n})}=0$ implies
\be
    \sum_n\Qb_{({\rm n})}=0,
    \qquad
    \sum_n\delta Q_{({\rm n})}=0,
    \qquad
    \sum_n f_{({\rm n})}=0.
\label{eq:Qsumzero}
\ee
For the special case of two interacting fluids, this reduces to $\Qb_{(1)}=-\Qb_{(2)}\,,
   \;\;
    \delta Q_{(1)}=-\delta Q_{(2)} \;\;\text{and} \;\;
    f_{(1)}=-f_{(2)}\, .
$
We have now decomposed $Q^\mu_{({\rm n})}$ into $\Qb_{({\rm n})}$, $\delta Q_{({\rm n})}$ and $f_{({\rm n})}$.
This sets all the rules for computing the relevant quantities and equations that describe the evolution of both the background and perturbations with any number of interacting components.

In the background, the continuity equation for each interacting component becomes
\be
    \rhob_{({\rm n})}'
    +
    3\mH(1+w_{({\rm n})})\rhob_{({\rm n})}
    =
    a\Qb_{({\rm n})} \,.
\label{eq:rho_background_interacting}
\ee
For the squared physical sound speed $c_{{\rm s},({\rm n})}^2$ and adiabatic sound speed \eqref{eq:soundspeed_adiabatic},
we get the relations
\begin{align}
\label{eq:cs_ca_interacting}
c_{{\rm a},({\rm n})}^2
&=
w_{({\rm n})}
+
\frac{w_{({\rm n})}'}
{-3\mH\left(1+w_{({\rm n})}\right)
+a\Qb_{({\rm n})}/\rhob_{({\rm n})}}
=
w_{({\rm n})}
+
\rhob_{({\rm n})}
\frac{w_{({\rm n})}'}{\rhob_{({\rm n})}'}\,,
\\[0.5em]
c_{{\rm s},({\rm n})}^2-c_{{\rm a},({\rm n})}^2
&=
c_{{\rm s},({\rm n})}^2-w_{({\rm n})}
-
\frac{w_{({\rm n})}'}
{-3\mH\left(1+w_{({\rm n})}\right)
+a\Qb_{({\rm n})}/\rhob_{({\rm n})}} \,.
\end{align}
With the conventions established above and the additional four-velocity source components \eqref{eq:Qcomponents},
the perturbed continuity and Euler equations for each component become%
\footnote{In the case of baryons, the Euler equation~\eqref{eq:theta_interacting} must again be supplemented with the Thomson-scattering term. Again, since $w_{\rm (b)}\simeq c_{{\rm a},({\rm b})}^2\simeq0$  the interaction terms proportional to these quantities vanish, while those proportional to $c_{{\rm s},({\rm b})}^2$ do not. Whether the latter should be neglected, as is usually done for the analogous terms in the non-interacting case, is not clearly addressed in the literature since although they are extremely small, in principle all of them should be consistently taken into account (see \cite{2020Univ....6....6P}, for example).}~\cite{Wang:2016lxa,Valiviita:2008iv}
\begin{align}
\delta_{({\rm n})}' ={}&
3\mathcal{H}\bigl(w_{({\rm n})}-c_{{\rm s},({\rm n})}^2\bigr)\delta_{({\rm n})}-(1+w_{({\rm n})})\theta_{({\rm n})}+3(1+w_{({\rm n})})\Phi'-9\mathcal{H}^2\bigl(c_{{\rm s},({\rm n})}^2-c_{{\rm a},({\rm n})}^2\bigr)(1+w_{({\rm n})})\frac{\theta_{({\rm n})}}{k^2}\nonumber\\
&+\frac{a}{\rho_{({\rm n})}}\delta Q_{({\rm n})}+\frac{a \Qb_{({\rm n})}}{\rho_{({\rm n})}}\left[\Psi-\delta_{({\rm n})}+3\mathcal{H}\bigl(c_{{\rm s},({\rm n})}^2-c_{{\rm a},({\rm n})}^2\bigr)\frac{\theta_{({\rm n})}}{k^2}\right] \,,
\label{eq:delta_interacting}
\\[1ex]
\theta_{({\rm n})}' ={}&
-\mathcal{H}\bigl(1-3c_{{\rm s},({\rm n})}^2\bigr)\theta_{({\rm n})}+\frac{k^2c_{{\rm s},({\rm n})}^2}{1+w_{({\rm n})}}\delta_{({\rm n})}+k^2\Psi\nonumber\\
&+\frac{a \Qb_{({\rm n})}}{(1+w_{({\rm n})})\rho_{({\rm n})}}\left[\hat{\theta}-\bigl(1+c_{{\rm s},({\rm n})}^2\bigr)\theta_{({\rm n})}\right]+\frac{ak^2}{(1+w_{({\rm n})})\rho_{({\rm n})}}f_{({\rm n})} \,.
\label{eq:theta_interacting}
\end{align}
The reference frame velocity is $\hat\theta$, as explained around equation \eqref{eq:fourvelocityref}.
With our chosen decomposition \eqref{eq:Qdecomp}, $\delta Q_{({\rm n})}$ and $f_{({\rm n})}$ enter the continuity and Euler equations separately,
while $\bar{Q}$ is a background quantity and enters both.

\subsubsection{Hubble rate, local expansion and the choice of reference frame}

Consider an interaction whose background coupling can be written as $Q=\mathcal{I}H$, where $\mathcal{I}$ is any scalar and $H$ is the Hubble rate.
A frequently adopted phenomenological prescription is to treat $H$ as unperturbed, which leads to $\delta Q=H\,\delta\mathcal{I}$.
At the homogeneous level, the Hubble rate is unambiguous.
In the perturbations, however, $H$ alone does not define a unique local covariant quantity, so a covariant completion is required.
Thus, we must specify a timelike congruence $u^\mu$, which at the linear level is nothing more than specifying a velocity reference frame, and replace the Hubble function by the expansion scalar $\Theta\equiv\nabla_\mu u^\mu$.
Therefore, the phenomenological choice of not perturbing $H$ leads to an incomplete covariant specification of the interaction.
Although often sufficient in the past, neglecting the perturbation of the local expansion may affect current and future precision observations.

Following the ambition of \SymBoltz to solve the complete Einstein--Boltzmann system with no approximation schemes, we include the fully covariant formulation here.
In the background ${\Theta}=\nabla_\mu u^\mu$ is simply $3H$, while up to first order it is~\cite{Gavela:2010tm,vonMarttens:2018iav}
\begin{equation}
    \Theta=\nabla_\mu u^\mu = \partial_\mu u^\mu + \Gamma^\mu_{\mu\lambda} u ^\lambda= \frac{3}{a} \left[ \mathcal{H}(1-\Psi) - \Phi' + \frac{\theta}{3} \right] \;.
\end{equation}
In the decomposition \eqref{eq:Qdecomp} of $Q^\mu$,
we introduced the four-vector $\hat{u}^\mu$ of the reference frame in which the energy transfer is measured
without defining exactly what that frame would be.
In this work we adopt the total energy frame as the reference frame.
It is constructed from the velocities of all cosmological components weighted by their respective inertial densities,
which provides a natural and general prescription for the frame in which the energy transfer is measured.
In our notation that means: 
\begin{equation}
\hat{\Theta}\equiv\Theta_{\rm tot}=\nabla_\mu \hat{u}^\mu \quad\implies\quad \hat{v}^i=\frac{\sum_{({\rm n})} (1+w_{({\rm n})})\bar\rho_{({\rm n})} v^i_{({\rm n})}}{\sum_{({\rm n})} (1+w_{({\rm n})})\bar\rho_{({\rm n})}}, \quad\text{i.e.}\quad \hat{\theta}=\frac{\sum_{({\rm n})} (1+w_{({\rm n})})\bar\rho_{({\rm n})} \theta_{({\rm n})}}{\sum_{({\rm n})} (1+w_{({\rm n})})\bar\rho_{({\rm n})}} 
\label{eq:total_vel}
\end{equation}
We put a hat $\;\hat{}\;$ on variables related to the total weighted velocity from now on.
Other choices for the reference frame are possible, in fact any other choice of the congruence is perfectly acceptable if used consistently.
Given any choice, we end up with
\begin{equation}
     \Bar{Q} = \frac{\Bar{\mathcal{I}}\Bar{\Theta}}{3},
    \qquad
    \delta Q
    =
    \frac{\Bar{\Theta}}{3}\,\delta \mathcal{I}
    +
    \frac{\mathcal{\Bar{I}}}{3}\,\delta\Theta ,
\end{equation}
where $\bar{\Theta}$ and $\delta\Theta$ are respectively the background expansion and its linear perturbation associated with the chosen congruence. In this work, we use the total energy frame defined above.

\section{Implementation in SymBoltz}
\label{sec:implementation}

In this section we show how interacting cosmological models can be implemented in \SymBoltz.
We want to show that the code for implementing a new interaction in \SymBoltz is considerably more direct than in previous works or Einstein--Boltzmann codes,
no matter how elaborate it may be or even if the interaction modifies several fluids or depends non-linearly on the densities.
Only three functions have to be implemented, following the decomposition explained in Section~\ref{subsec:cov_couplings}. 

Once the user has expressions $\bar{Q}_{\rm(n)(m)}$, $\delta Q_{\rm(n)(m)}$ and $f_{\rm(n)(m)}$ for the interaction between one or more fluid pairs $\{{\rm(n),(m)}\}$, they are ready to be implemented in \SymBoltz.
We adopt the following sign convention for $Q^\mu$ (and for $\bar{Q}$, $\delta Q$ and $f$) to satisfy condition \eqref{eq:Qsumzero}:
\be
Q^\mu_{\rm (b)} = -Q^\mu_{\rm bc}+Q^\mu_{{\rm b}\Lambda}, \qquad
Q^\mu_{\rm (c)} = Q^\mu_{\rm bc}+Q^\mu_{{\rm c}\Lambda}, \qquad
Q^\mu_{(\Lambda)} = -Q^\mu_{{\rm b}\Lambda}-Q^\mu_{{\rm c}\Lambda}.
\label{eq:interaction-sign-convention}
\ee

We modify the minimal \LCDM{} example in the code documentation,
where every variable, parameter and equation of the entire standard model is specified in around 300 lines of code.
This example is straightforward to modify with arbitrary interaction equations.
To not duplicate the entire model for every interaction,
we wrap this in a function \verb|interacting_model(Qbc, δQbc, fQbc, QbΛ, ...)|%
\footnote{Our full model definition is found at \url{https://hersle.github.io/SymBoltz.jl/stable/interacting/}.}
that takes arbitrary interaction expressions as input.
This function can now be used to create models with arbitrary pairwise interactions by passing symbolic expressions.
We first create a non-interacting model \texttt{M} to get a handle to all its symbolic expressions,
and then create several \texttt{interacting\_model()} instances with arbitrary interaction expressions.
For example, to conveniently construct the interacting model~\eqref{eq:de_dm_linear}:
\begin{snugshade}
\begin{Verbatim}[commandchars=\\\{\},fontsize=\small,breaklines,xleftmargin=5pt,xrightmargin=5pt]
\PY{c}{\PYZsh{} Build non\PYZhy{}interacting base model to access its symbolic variables}
\PY{n}{M}\PY{+w}{ }\PY{o}{=}\PY{+w}{ }\PY{n}{interacting\PYZus{}model}\PY{p}{(}\PY{p}{)}
\PY{n+nd}{@unpack}\PY{+w}{ }\PY{n}{a}\PY{p}{,}\PY{+w}{ }\PY{n}{ℋ}\PY{p}{,}\PY{+w}{ }\PY{n}{ρc}\PY{p}{,}\PY{+w}{ }\PY{n}{ρΛ}\PY{p}{,}\PY{+w}{ }\PY{n}{δc}\PY{p}{,}\PY{+w}{ }\PY{n}{δΛ}\PY{p}{,}\PY{+w}{ }\PY{n}{θ}\PY{p}{,}\PY{+w}{ }\PY{n}{θc}\PY{p}{,}\PY{+w}{ }\PY{n}{θΛ}\PY{p}{,}\PY{+w}{ }\PY{n}{Ψ}\PY{p}{,}\PY{+w}{ }\PY{n}{Φ}\PY{p}{,}\PY{+w}{ }\PY{n}{k}\PY{p}{,}\PY{+w}{ }\PY{n}{αcΛ}\PY{+w}{ }\PY{o}{=}\PY{+w}{ }\PY{n}{M}

\PY{c}{\PYZsh{} Create interacting model instances}
\PY{n}{M1}\PY{+w}{ }\PY{o}{=}\PY{+w}{ }\PY{n}{interacting\PYZus{}model}\PY{p}{(}
\PY{+w}{    }\PY{n}{name}\PY{+w}{ }\PY{o}{=}\PY{+w}{ }\PY{l+s+ss}{:ζCDM}\PY{p}{,}
\PY{+w}{    }\PY{n}{QcΛ}\PY{+w}{ }\PY{o}{=}\PY{+w}{ }\PY{n}{αcΛ}\PY{+w}{ }\PY{o}{*}\PY{+w}{ }\PY{n}{ℋ}\PY{+w}{ }\PY{o}{*}\PY{+w}{ }\PY{n}{ρΛ}\PY{+w}{ }\PY{o}{/}\PY{+w}{ }\PY{n}{a}\PY{p}{,}
\PY{+w}{    }\PY{n}{δQcΛ}\PY{+w}{ }\PY{o}{=}\PY{+w}{ }\PY{n}{αcΛ}\PY{+w}{ }\PY{o}{/}\PY{+w}{ }\PY{n}{a}\PY{+w}{ }\PY{o}{*}\PY{+w}{ }\PY{p}{(}\PY{n}{ℋ}\PY{+w}{ }\PY{o}{*}\PY{+w}{ }\PY{p}{(}\PY{n}{ρΛ}\PY{o}{*}\PY{n}{δΛ}\PY{+w}{ }\PY{o}{\PYZhy{}}\PY{+w}{ }\PY{n}{ρΛ}\PY{o}{*}\PY{n}{Ψ}\PY{p}{)}\PY{+w}{ }\PY{o}{+}\PY{+w}{ }\PY{n}{ρΛ}\PY{+w}{ }\PY{o}{*}\PY{+w}{ }\PY{p}{(}\PY{n}{θ}\PY{o}{/}\PY{l+m+mi}{3}\PY{+w}{ }\PY{o}{\PYZhy{}}\PY{+w}{ }\PY{n}{Dτ}\PY{p}{(}\PY{n}{Φ}\PY{p}{)}\PY{p}{)}\PY{p}{)}\PY{p}{,}
\PY{+w}{    }\PY{n}{fQcΛ}\PY{+w}{ }\PY{o}{=}\PY{+w}{ }\PY{n}{αcΛ}\PY{+w}{ }\PY{o}{*}\PY{+w}{ }\PY{n}{ℋ}\PY{+w}{ }\PY{o}{*}\PY{+w}{ }\PY{n}{ρΛ}\PY{+w}{ }\PY{o}{/}\PY{+w}{ }\PY{p}{(}\PY{n}{a}\PY{o}{*}\PY{n}{k}\PY{o}{\PYZca{}}\PY{l+m+mi}{2}\PY{p}{)}\PY{+w}{ }\PY{o}{*}\PY{+w}{ }\PY{p}{(}\PY{n}{θc}\PY{+w}{ }\PY{o}{\PYZhy{}}\PY{+w}{ }\PY{n}{θ}\PY{p}{)}\PY{p}{,}
\PY{p}{)}
\end{Verbatim}
\end{snugshade}
Other interacting models are constructed similarly.
This makes it easy to create many different interacting models.

\section{Validation}
\label{sec:validation}

We now compare some simple examples of interacting models in \SymBoltz with modified versions of the well-established \CLASS code from earlier works.
Our goal is to demonstrate that \SymBoltz reproduces the evolution of the background and perturbations with a similar level of precision as \CLASS.
First we consider standard cosmological models without interactions, then pure momentum transfer models, and finally models with both energy and momentum exchange
(see Ref.~\cite{Wang:2024vmw} for a recent review on interactions).
This allows us to separately check the standard non-interacting evolution, modifications affecting only the perturbations, and cases in which the interaction also changes the background evolution.

We compare background densities, the Hubble function, matter power spectra, and angular CMB TT and EE power spectra.
We fix the reduced Hubble constant to $h=0.70$, the dark matter density today to $\Omega_{\rm c}=0.25$, the baryon density today to $\Omega_{\rm b}=0.05$, the helium fraction to $Y_{\rm He}=0.25$, the amplitude of scalar perturbations to $A_{\rm s}=2\;\cdot10^{-9}$, the scalar spectral index to $n_{\rm s}=0.94$, and use two massless neutrinos and one massive neutrino with total mass $\sum_i m_\nu =0.02\;\text{eV}$.
Other parameters are fixed to default values.
The results of the comparison are similar for other parameter values.

\subsection{Reference cases: no interactions}

\begin{figure}[H]
    \centering
    \vspace{-0.5cm}
    \includegraphics[width=0.9\textwidth]{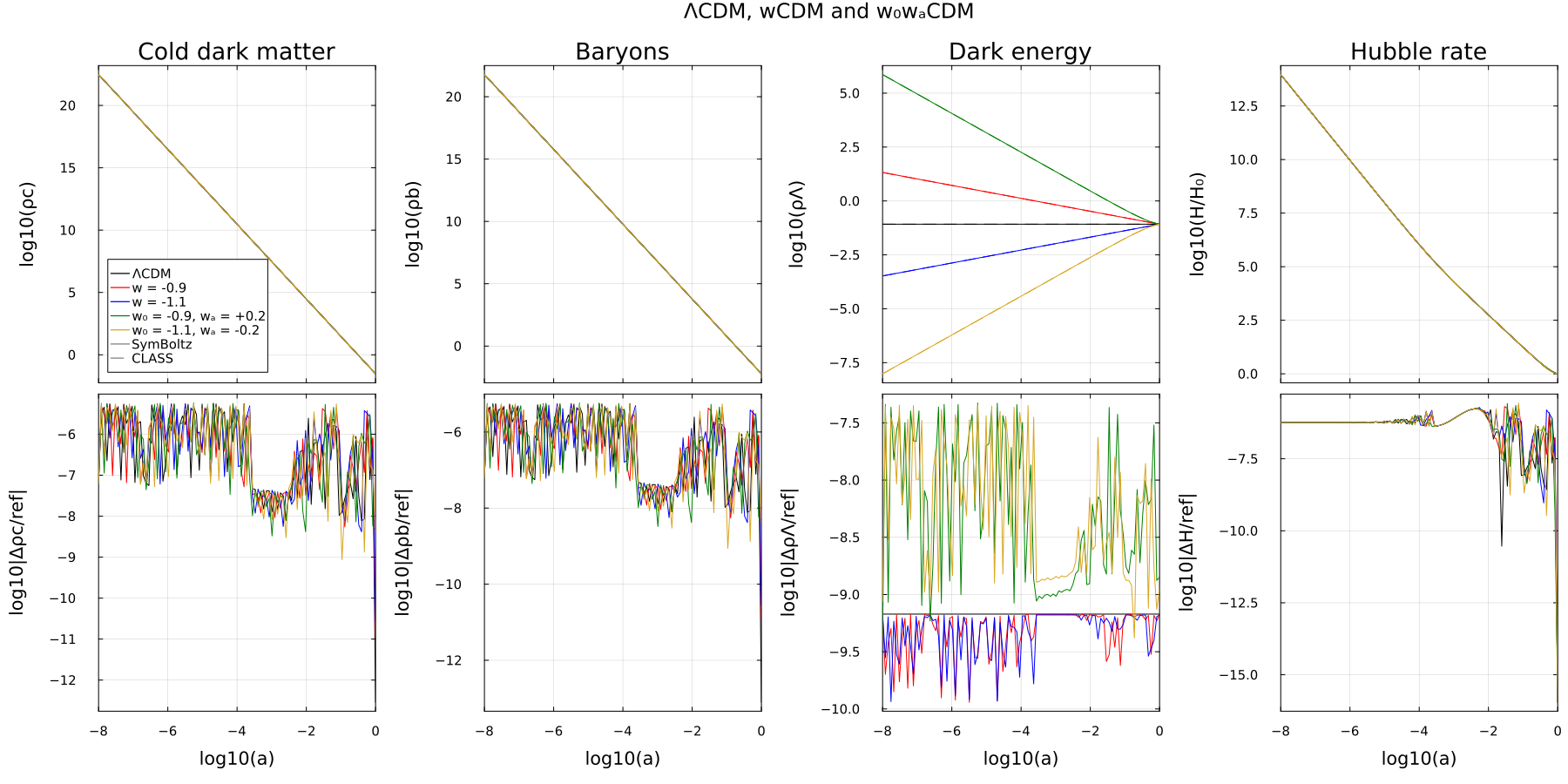}

    \caption{Background comparison between \CLASS and \SymBoltz for $\Lambda$CDM (black), $w$CDM with $w=-0.9$ (red) and $w=-1.1$ (blue), and $w_0w_a$CDM with $(w_0,w_a)=(-0.9,+0.2)$ (green) and $(w_0,w_a)=(-1.1,-0.2)$ (gold). The upper panels show the background evolution, while the lower panels show the relative difference, calculated as $\log_{10}\left|\left(X_{\rm \SymBoltz}-X_{\rm \CLASS}\right)/X_{\rm \CLASS}\right|$.}
    \label{fig:ref_background_CLASS_vs_SymBoltz}
\end{figure}

\begin{figure}[H]
    \centering
    \vspace{-0.5cm}
    \includegraphics[height=5.9cm]{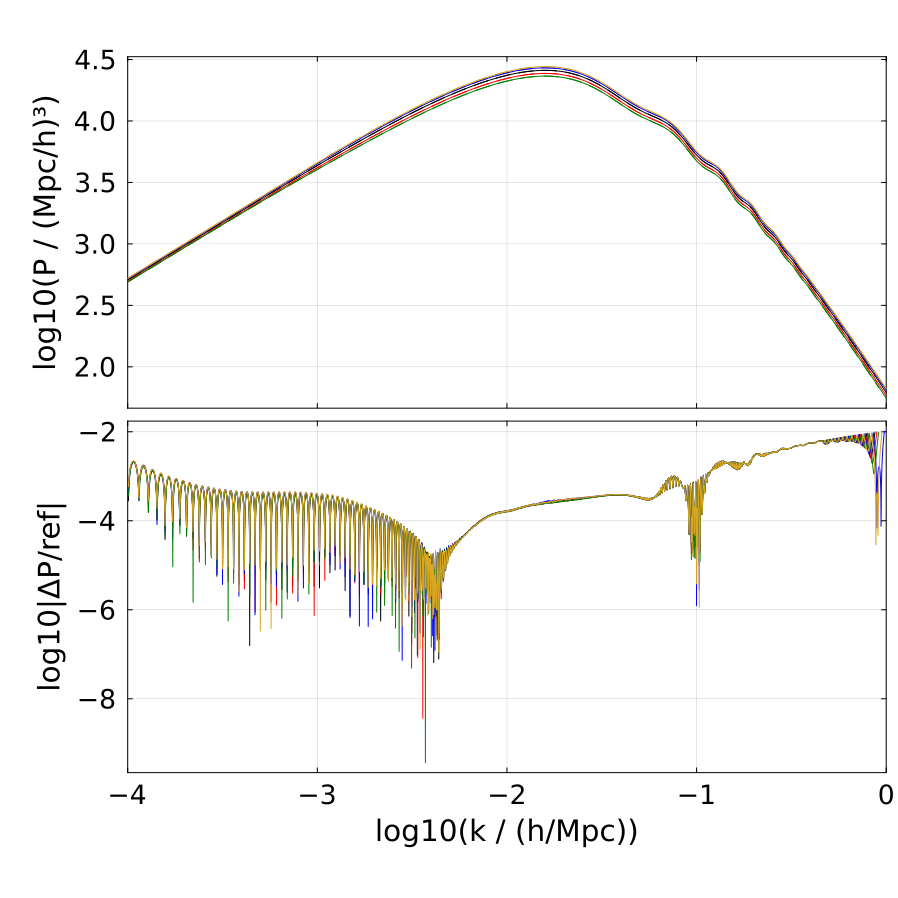}
    \includegraphics[height=5.9cm]{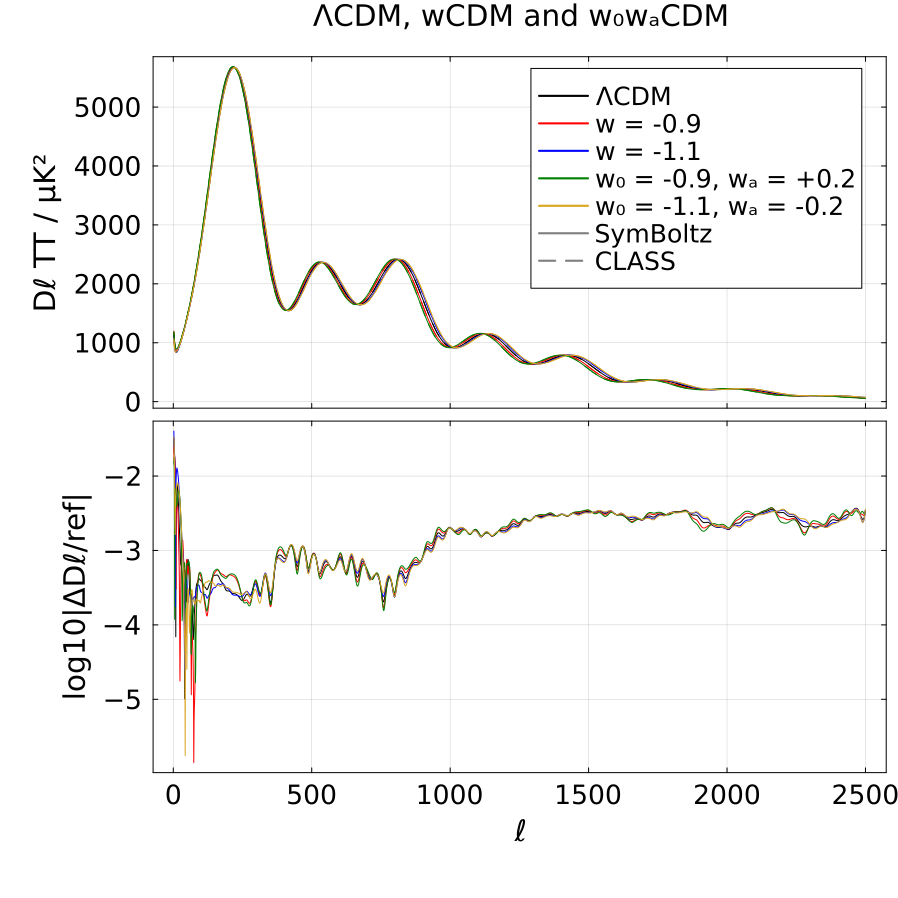}
    \includegraphics[height=5.9cm]{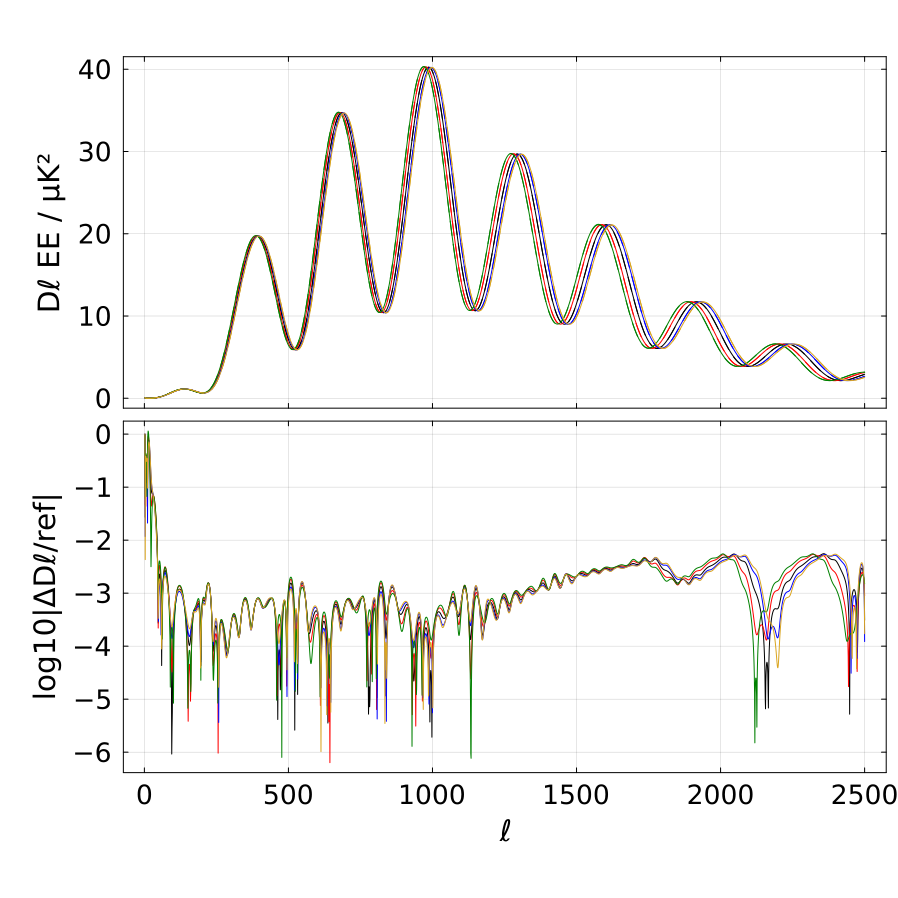}
    \caption{Comparison of matter, CMB TT and CMB EE power spectra (left to right) between \CLASS and \SymBoltz for $\Lambda$CDM (black), $w$CDM with $w=-0.9$ (red) and $w=-1.1$ (blue), and $w_0w_a$CDM with $(w_0,w_a)=(-0.9,+0.2)$ (green) and $(w_0,w_a)=(-1.1,-0.2)$ (gold). The lower panels show the logarithm of the relative difference $\log_{10}\left|\left(X_{\rm \SymBoltz}-X_{\rm \CLASS}\right)/X_{\rm \CLASS}\right|$.}
    \label{fig:ref_pk_cmb_CLASS_vs_SymBoltz}
\end{figure}

As a reference case, we first compare \SymBoltz with \CLASS for the non-interacting $\Lambda$CDM, $w$CDM and $w_0w_a$CDM models,
the latter of which includes the Chevallier--Polarski--Linder (CPL) dark energy parametrization~\cite{2001IJMPD..10..213C,2003PhRvL..90i1301L}.
The first and most important result is that \SymBoltz reproduces accurate results for these standard models in the absence of interactions.
Figures~\ref{fig:ref_background_CLASS_vs_SymBoltz} and~\ref{fig:ref_pk_cmb_CLASS_vs_SymBoltz} compare one $\Lambda$CDM model, two representative $w$CDM models and two representative $w_0w_a$CDM models.
In the background, the cold dark matter density $\rho_{\rm c}$, baryon density $\rho_{\rm b}$, dark energy/cosmological constant density $\rho_\Lambda$ and Hubble rate $H(a)$ from \SymBoltz follow the corresponding \CLASS curves closely in all five cases.
The curves lie on top of each other and the relative difference with respect to \CLASS remains well below $0.001\%$.
The matter power spectrum agrees below the $1\%$ level for all $k\in[10^{-4}, 1]\; h/\text{Mpc}$.
Finally, the CMB power spectra also agree well within $1\%$ except at very low multipoles in the EE spectrum, where the spectrum approaches zero.
The resulting spike in the relative difference is therefore caused by the normalization to a vanishingly small \CLASS value, rather than by a significant absolute discrepancy between the two codes.
This is also the domain where cosmic variance dominates the uncertainty anyway.

\subsection{Pure momentum transfer}

    Now we consider interacting models implemented in \SymBoltz.
    We begin with the simplest case one can imagine: the trilogy of momentum transfer models.
    Here a coupling between the velocities of two species appears in their Euler equations (hence the name),
    while the perturbed continuity equation is unchanged in form (but not solution) and the entire background evolution is unchanged.
    One of the first appearances of these models dates back to Ref.~\cite{Simpson:2010vh}, but a plethora of models around the original one have emerged~\cite{Pourtsidou:2013nha,Boehmer:2015sha,Pourtsidou:2016ico,Linton:2017ged,Chamings:2019kcl,Amendola:2020ldb,BeltranJimenez:2020qdu,Figueruelo:2021elm,Linton:2021cgd,BeltranJimenez:2021wbq,Liu:2023mwx,Cruickshank:2025chm,Aoki:2025bmj,Jensko:2026taf,BeltranJimenez:2026ymd}.

    Here we consider three example models that we implement in \SymBoltz:
    the $\alpha$CDM model~\cite{Asghari:2019qld},
    the $\beta$CDM model~\cite{BeltranJimenez:2020iyx} and
    the $\gamma$CDM model~\cite{Figueruelo:2026wzk}.
    For these models we have access to corresponding modified versions of \CLASS,
    which we can then compare with \SymBoltz.
    
\subsubsection{Dark energy -- dark matter}
\begin{figure}[H]
    \centering
    \vspace{-0.5cm}
    \includegraphics[width=0.9\textwidth]{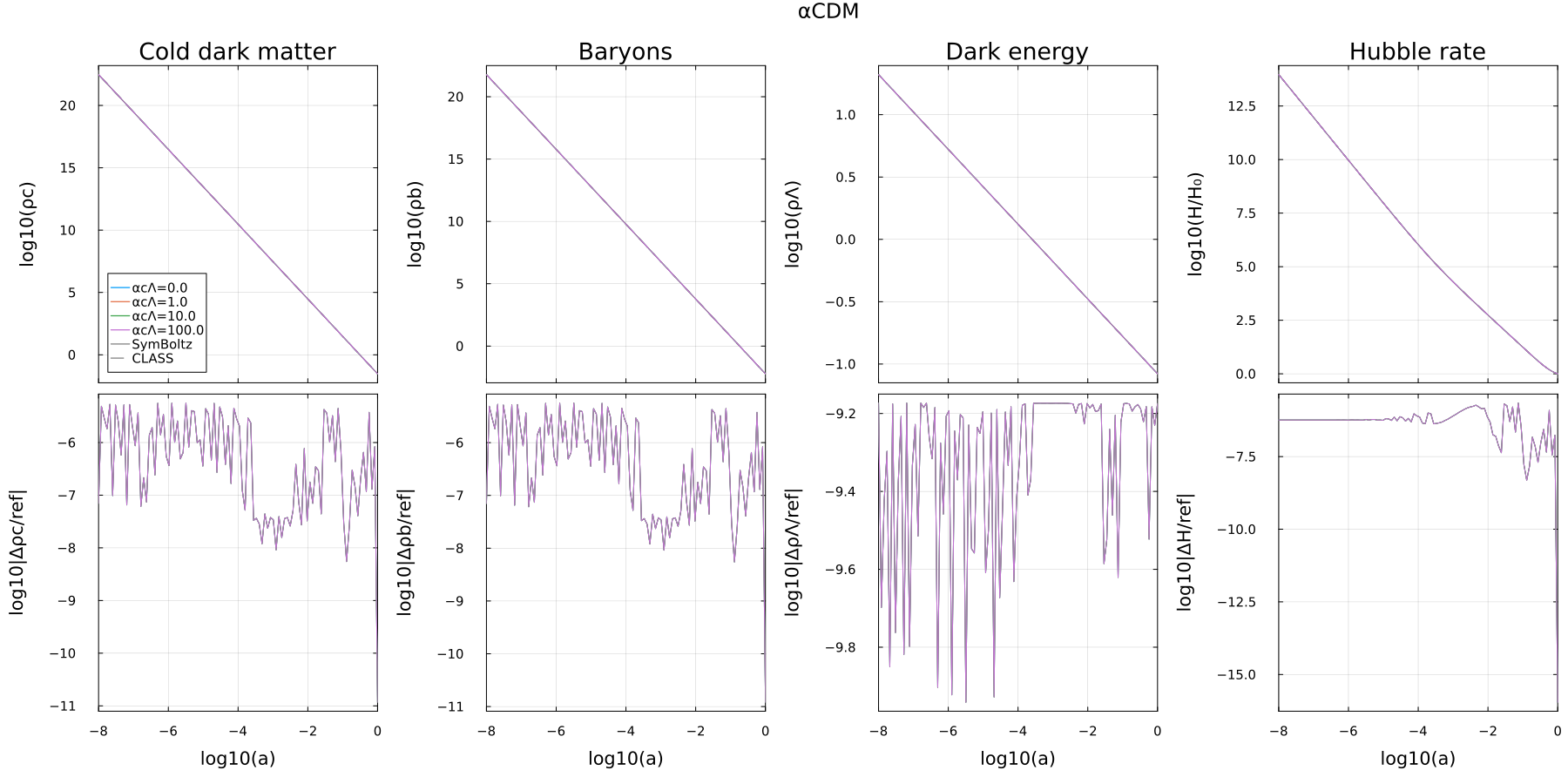}

    \caption{Background comparison between \CLASS and \SymBoltz as in Figure~\ref{fig:ref_background_CLASS_vs_SymBoltz}, but for the $\alpha$CDM model \eqref{eq:acdm_momentum}.}
    \label{fig:alphaCDM_background_CLASS_vs_SymBoltz}
\end{figure}

\begin{figure}[H]
    \centering
    \vspace{-0.5cm}
    \includegraphics[height=5.9cm]{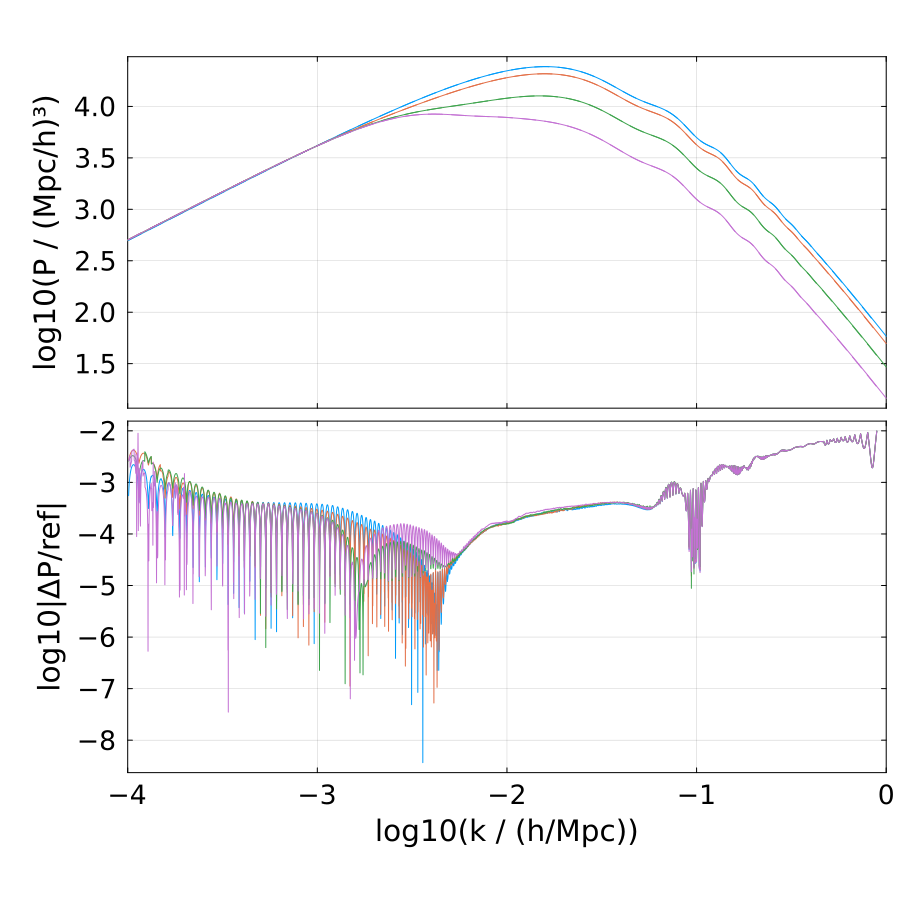}
    \includegraphics[height=5.9cm]{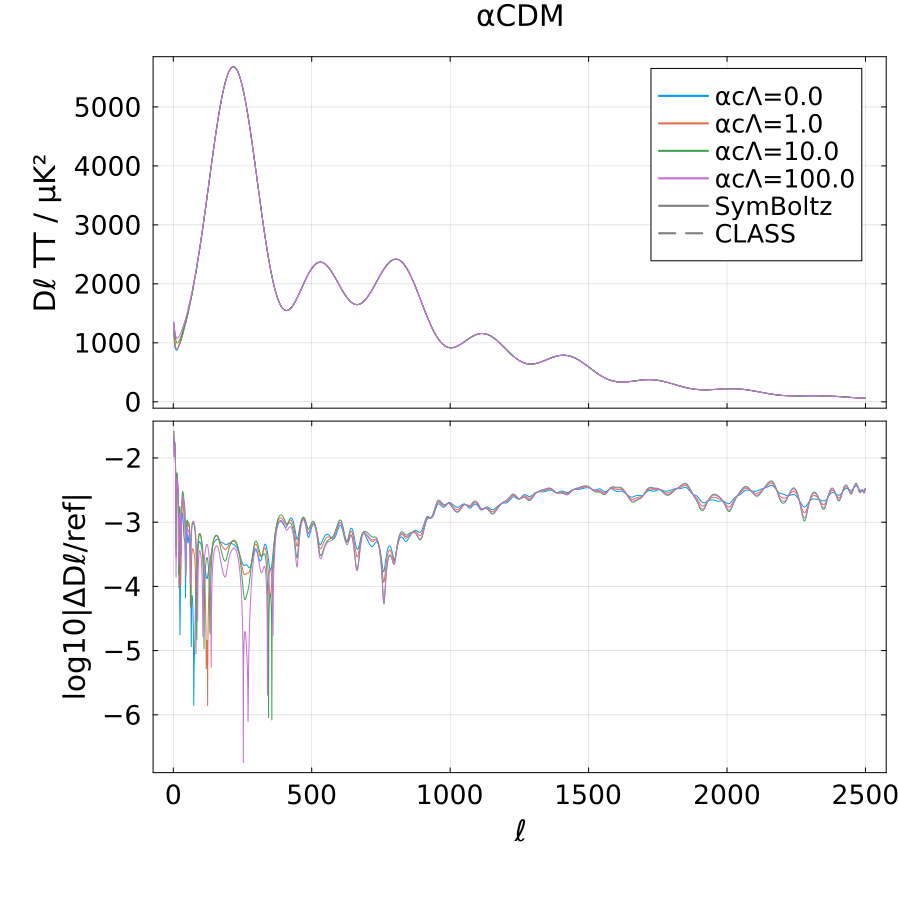}
    \includegraphics[height=5.9cm]{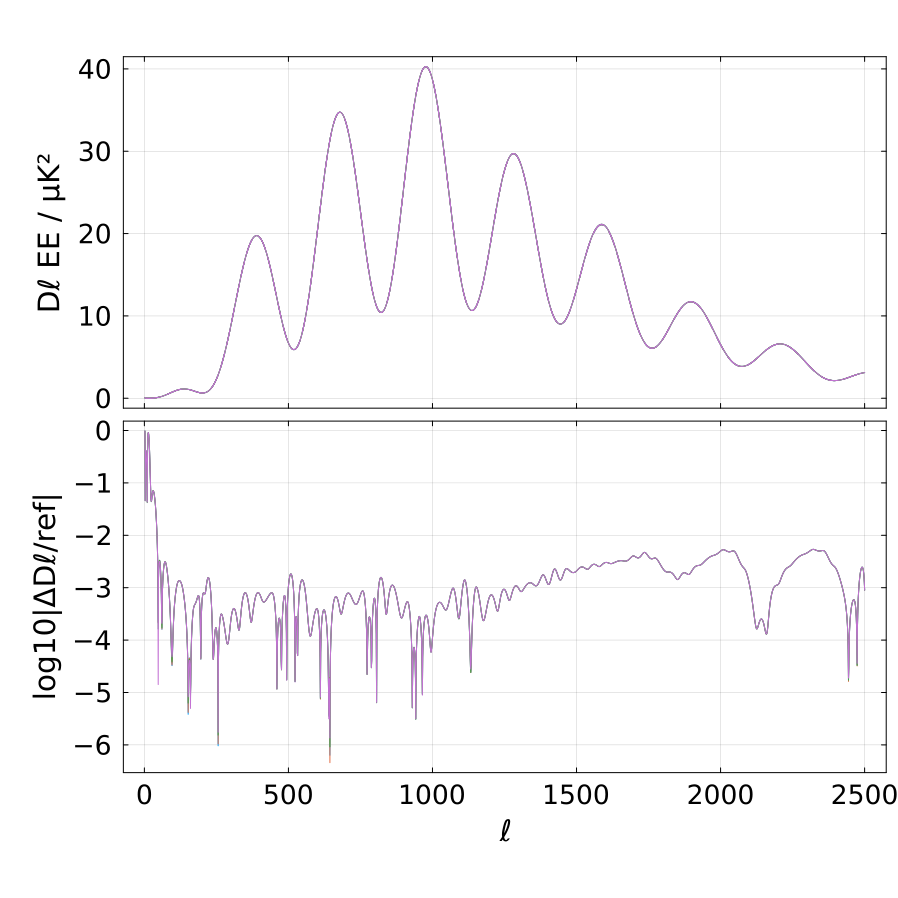}

    \caption{Power spectrum comparison between \CLASS and \SymBoltz as in Figure~\ref{fig:ref_pk_cmb_CLASS_vs_SymBoltz}, but for the $\alpha$CDM model \eqref{eq:acdm_momentum}.}
    \label{fig:alphaCDM_pk_cmb_CLASS_vs_SymBoltz}
\end{figure}

        A pure momentum interaction between dark matter and dark energy was first studied in Ref.~\cite{Asghari:2019qld} and more recently extended in Ref.~\cite{Figueruelo:2021elm}.
        The interaction is defined by $Q^\mu = \bar{\alpha}(u^\mu_{\Lambda} - u^\mu_{\rm c})$, and the decomposition of Section~\ref{subsec:cov_couplings} gives:%
        \footnote{In the code the coupling parameter always follows the notation $\alpha_{{\rm n m}}$, where $\rm n $ and $\rm m$ are pairwise interacting species. Here we simply use $\alpha$ for this interaction to ease notation (and $\beta$ and $\gamma$ in the other pure momentum cases), which corresponds to $\alpha\equiv\alpha{\rm c}{\Lambda}$ in \SymBoltz.} 
\be
\bar{Q}  = 0, \qquad
\delta Q = 0, \qquad
f        = \frac{\bar{\alpha}}{k^2} ( \theta_{\Lambda} - \theta_{\rm c})\qquad \text{with}\quad \bar{\alpha} =\alpha \frac{3 H_0^3}{8\pi G} = \alpha H_0 \rho_{\rm cr} .
\label{eq:acdm_momentum}
\ee
        This is implemented in \SymBoltz as described in Section~\ref{sec:implementation},
        and a corresponding modified version of \CLASS\footnote{\url{https://github.com/david-figuer/CLASS_momentum_transfer}} is available.
        Figures~\ref{fig:alphaCDM_background_CLASS_vs_SymBoltz} and \ref{fig:alphaCDM_pk_cmb_CLASS_vs_SymBoltz} again show the comparison between \SymBoltz and \CLASS.
        The background is not modified by this interaction,
        but we compare it again because the modified version of \CLASS is different from the base version of \CLASS used in the first non-interacting comparison.
        Both the background and the matter and CMB power spectra remain in similar agreement as in the non-interacting case, with similar $0.1\%$--$1\%$ differences in the power spectra.
        This holds over a large range of coupling parameters $\alpha$.
        Consequently, the \SymBoltz implementation demonstrates a very good match with previous implementations.

\subsubsection{Dark energy -- baryons}

\begin{figure}[H]
      \centering
        \vspace{+0.5cm}
       \includegraphics[height=5.9cm]{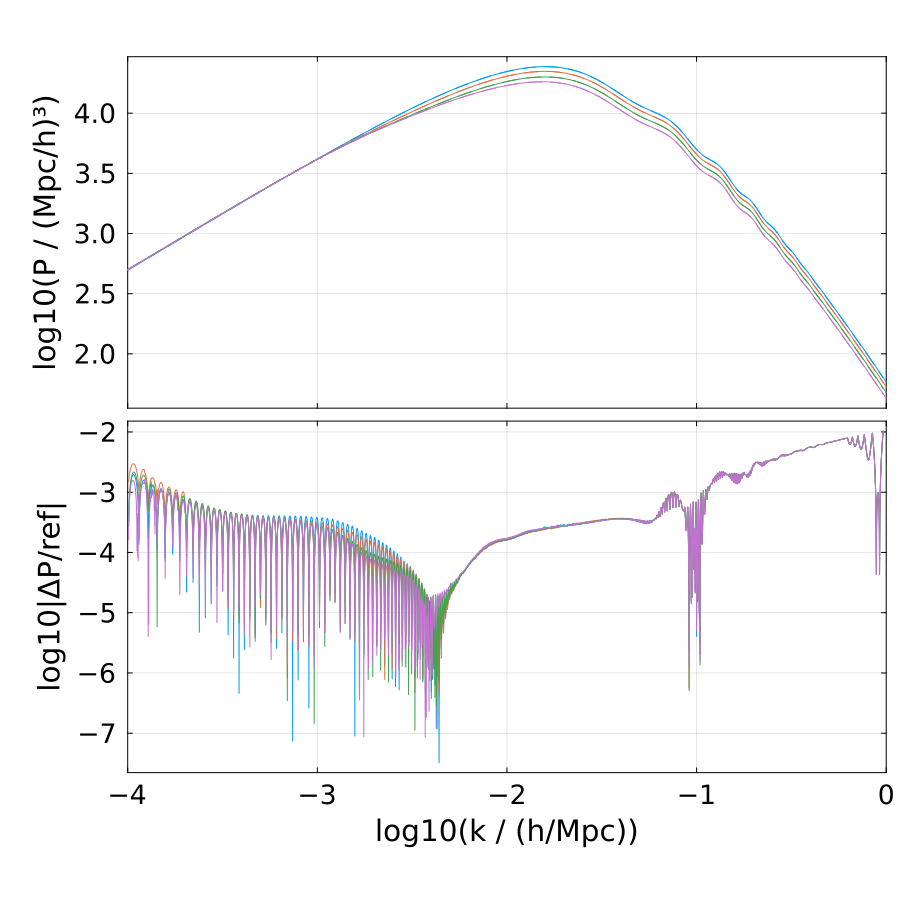}
    \includegraphics[height=5.9cm]{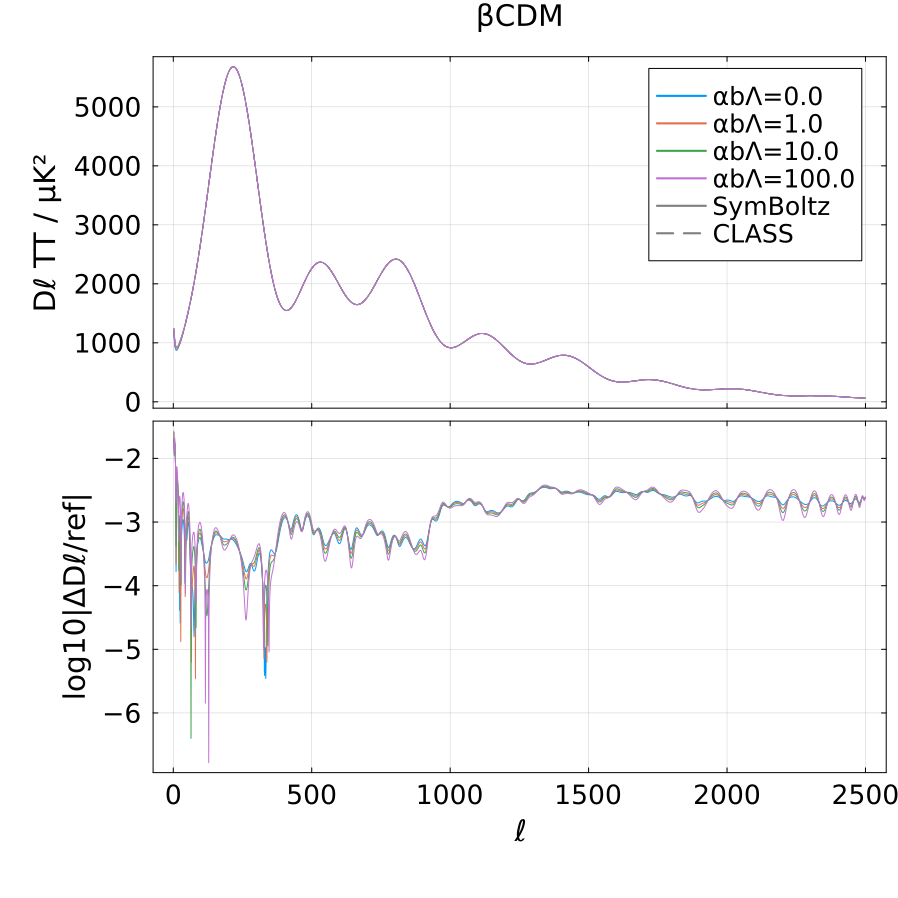}
     \includegraphics[height=5.9cm]{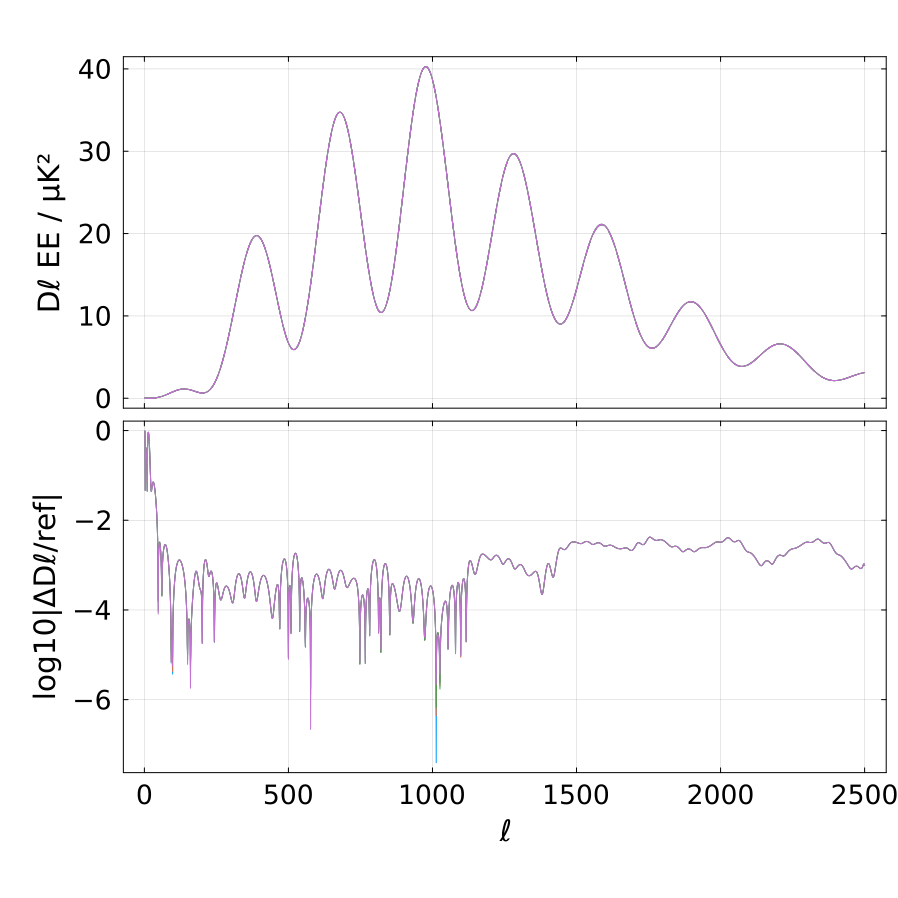}

    \caption{Power spectrum comparison between \CLASS and \SymBoltz as in Figure~\ref{fig:ref_pk_cmb_CLASS_vs_SymBoltz}, but for the $\beta$CDM model \eqref{eq:bcdm_momentum}. }
    \label{fig:betaCDM_CLASS_vs_SymBoltz}
\end{figure}

A pure momentum interaction between baryons and dark energy was first studied in~\cite{BeltranJimenez:2020iyx} inspired by Ref.~\cite{Vagnozzi:2019kvw}.
The interaction is defined by $Q^\mu = \bar{\beta}(u^\mu_{\Lambda} - u^\mu_{\rm b})$, and the decomposition gives:
\be
\bar{Q}  = 0, \qquad
\delta Q = 0, \qquad
f        = \frac{\bar{\beta}}{k^2} ( \theta_{\Lambda} - \theta_{\rm b}) \qquad \text{with}\quad \bar{\beta} =\beta \frac{3 H_0^3}{8\pi G} = \beta H_0 \rho_{\rm cr} .
\label{eq:bcdm_momentum}
\ee
        In Figure~\ref{fig:betaCDM_CLASS_vs_SymBoltz} we again compare the matter and CMB power spectra, while we omit the comparison of the background because it is not altered and we showed above that \SymBoltz and \CLASS are indistinguishable here.
        The matter and CMB power spectra again agree with less than $1\%$ difference.
        
\subsubsection{Dark matter -- baryons}

\begin{figure}[H]
      \centering
        \vspace{-0.2cm}
       \includegraphics[height=5.9cm]{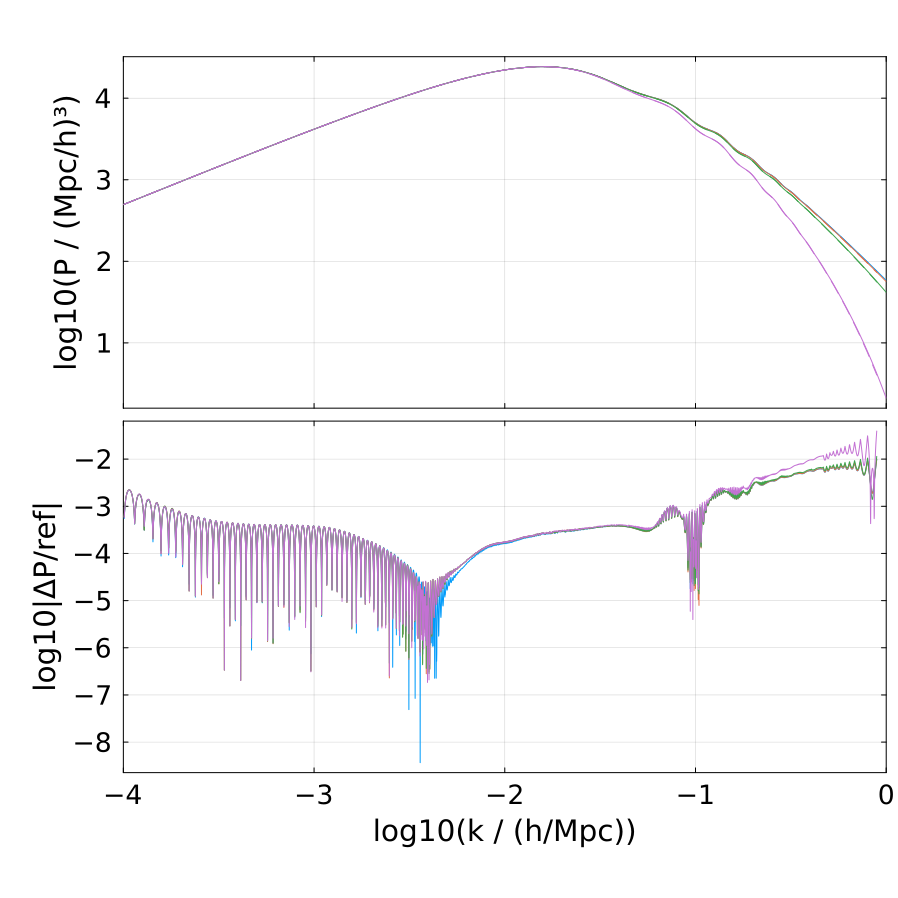}
    \includegraphics[height=5.9cm]{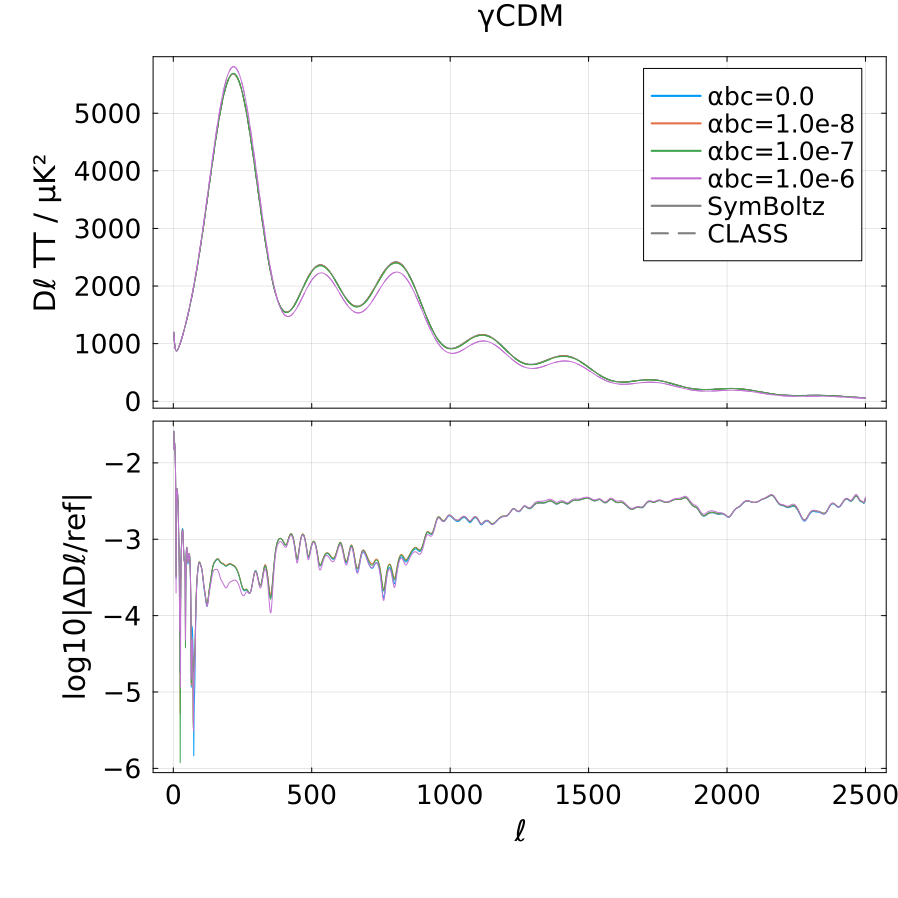}
    \includegraphics[height=5.9cm]{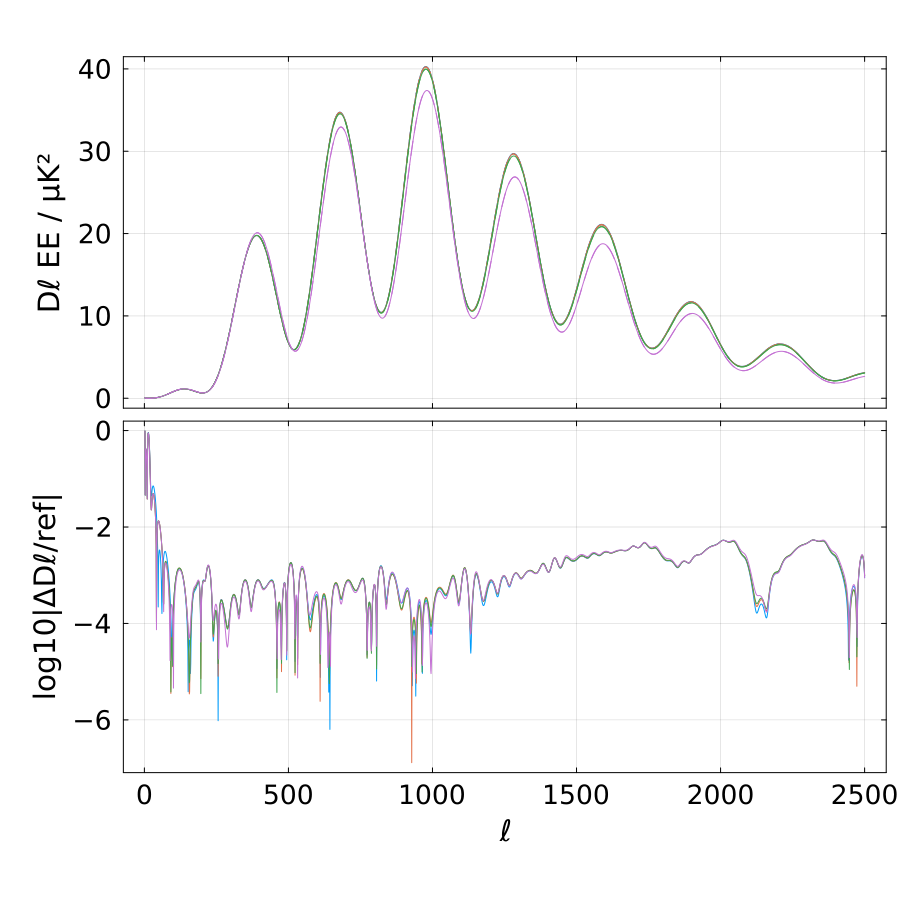}
    
    \caption{Power spectrum comparison between \CLASS and \SymBoltz as in Figure~\ref{fig:ref_pk_cmb_CLASS_vs_SymBoltz}, but for the $\gamma$CDM model \eqref{eq:gcdm_momentum}.}
    \label{fig:gammaCDM_CLASS_vs_SymBoltz}
\end{figure}

Finally, a pure momentum interaction between dark matter and baryons was first studied in Ref.~\cite{Figueruelo:2026wzk}.
The interaction is defined by $Q^\mu= \frac{\gamma}{\Omega_{\rm b} \rho_{cr,0}} (T^{\alpha\beta} u_\alpha^b u_\beta^b )^2 (u_{\rm b}^\mu - u_{\rm c}^\mu)$,
and the decomposition gives:
\be
\bar{Q}  =  0, \qquad
\delta Q =  0, \qquad
f        =  - \gamma \frac{ \Omega_{{\rm b},0} \rho_{cr,0}}{a^6 k^2} (\theta_{\rm c} - \theta_{\rm b}) .
\label{eq:gcdm_momentum}
\ee
We again omit the comparison of the unchanged background.
The matter power spectrum in Figure~\ref{fig:gammaCDM_CLASS_vs_SymBoltz} agrees well below the $1\%$ level for linear scales,
but some percent-level differences appear at smaller scales $k> 0.1\; h/\text{Mpc}$ for very high $\gamma$.
Here a non-linear prescription or $N$-body analysis is needed for accurate results anyway.
On the other hand, the CMB spectra continue to agree well within $1\%$ for different coupling values, similarly as before.

    \subsection{Dark energy -- dark matter: energy and momentum transfer}    

    Previous interactions involved only a pure momentum exchange and did not change the background evolution from the standard non-interacting $\Lambda$CDM, $w$CDM and $w_0w_a$CDM models.
    Now we extend the testing of the capabilities of \SymBoltz to scenarios in which energy is also transferred and the background evolution is changed.
    This will also serve as a test of the backwards integration capabilities of \SymBoltz version 2 explained in Section~\ref{sec:SB_v2}, since the interacting continuity equation \label{eq:rho_background_interacting} has no analytical solution for general $\Bar{Q}_{\rm (n)} \neq 0$ and must be solved using either the shooting method or backwards integration.
    As our example model, we consider a dark energy -- dark matter interaction with both energy and momentum transfer,
    first using a linear interaction term and then a non-linear one. Both cases have well-known analytical solutions for the background densities and Hubble function which we have implemented in the \CLASS versions. However, in \SymBoltz we have not used them. For the background in \SymBoltz we followed the general prescription for interactions, so a triad ($Q$, $\delta Q$, $f$) is introduced in each case, and then solved the background equations with the new backwards integration of \SymBoltz version 2. The comparison will show the capabilities of \SymBoltz to correctly solve the background dynamics with this new backwards integration possibility. 
\subsubsection{Linear coupling}

\begin{figure}[H]
    \centering
    \vspace{-0.5cm}
    \includegraphics[width=0.9\textwidth]{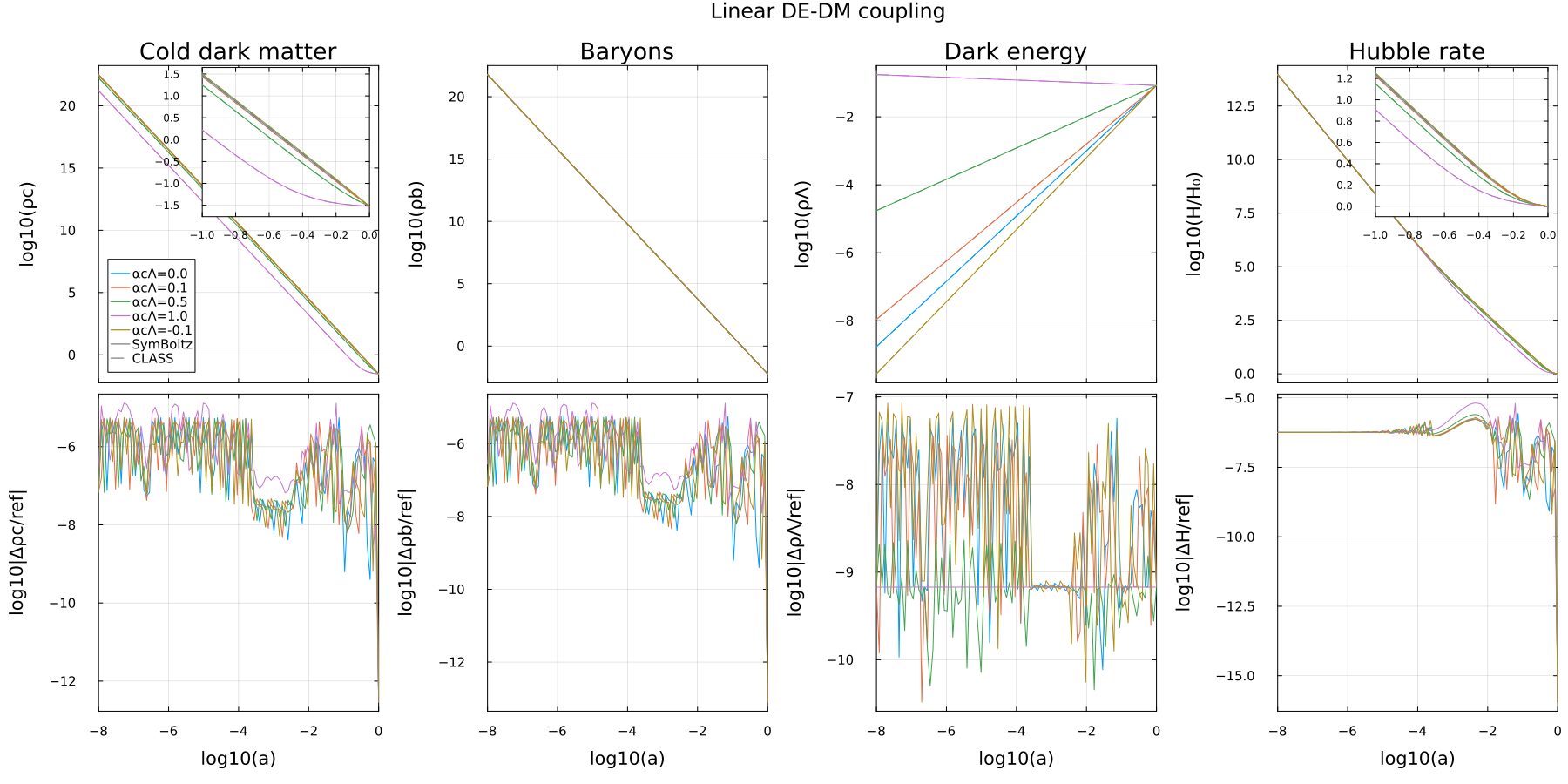}

    \caption{Background comparison between \CLASS and \SymBoltz as in Figure~\ref{fig:ref_background_CLASS_vs_SymBoltz}, but for the linear DE--DM interaction \eqref{eq:de_dm_linear}.}
    \label{fig:DE-DM-rhoDE-background}
\end{figure}

\begin{figure}[H]
    \centering
    \vspace{+0.5cm}
    \includegraphics[height=5.9cm]{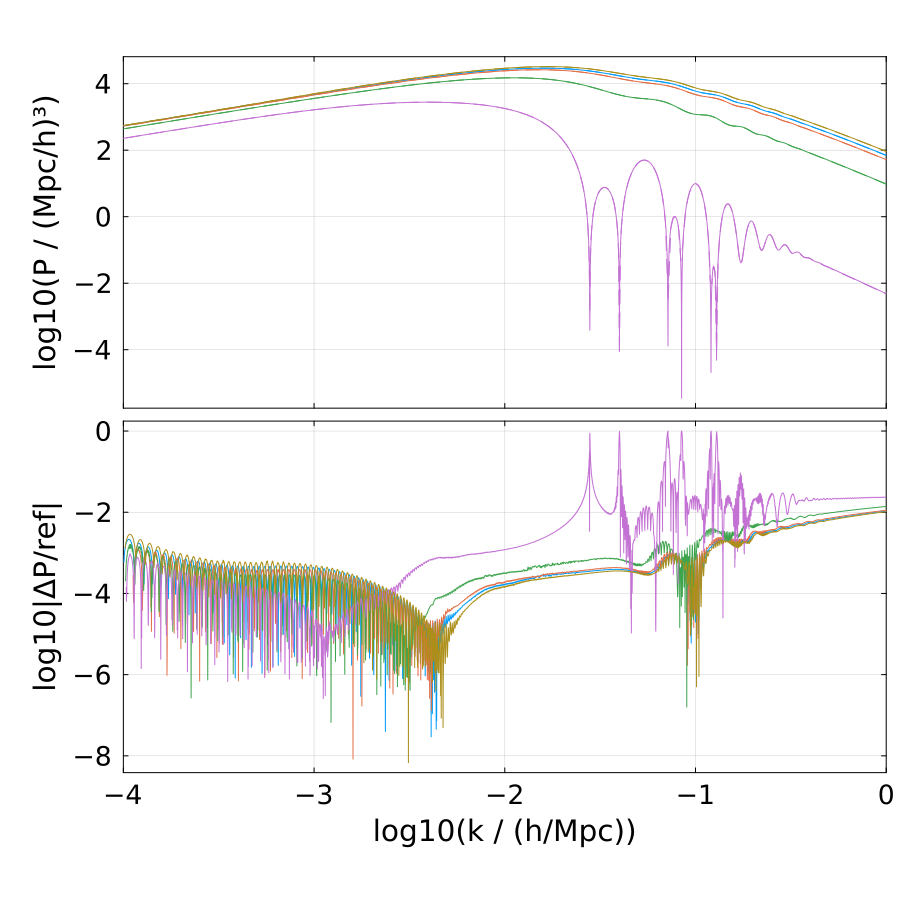}
    \includegraphics[height=5.9cm]{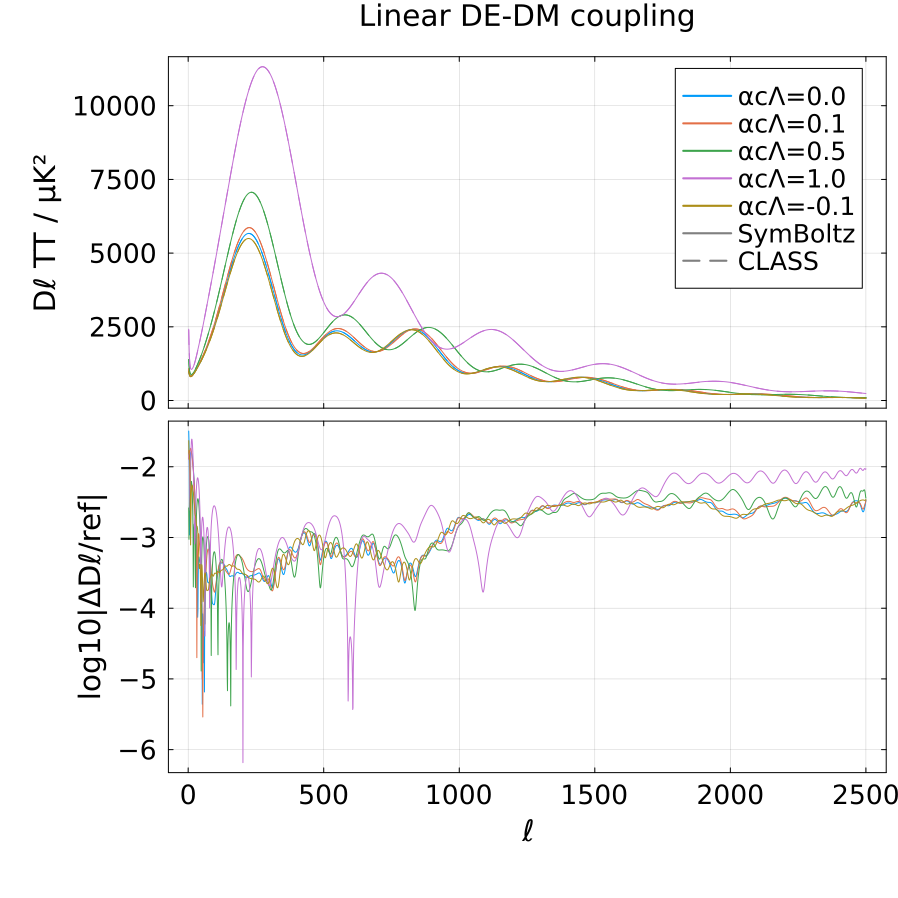}
    \includegraphics[height=5.9cm]{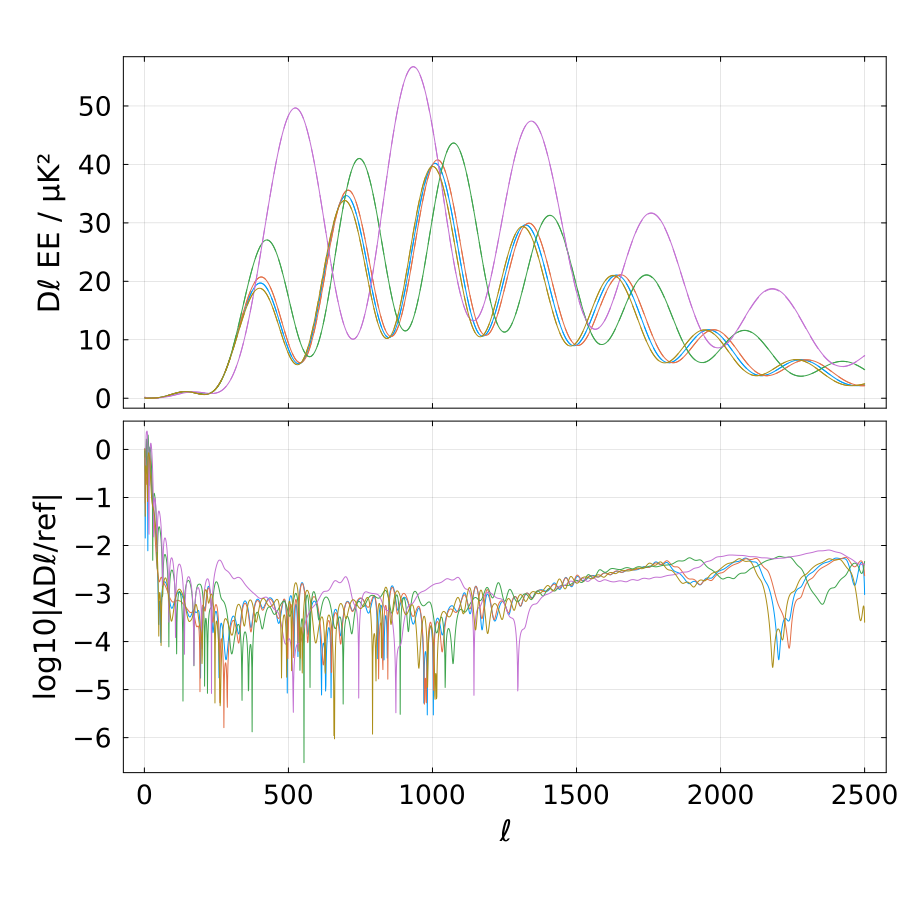}

    \caption{Power spectrum comparison between \CLASS and \SymBoltz as in Figure~\ref{fig:ref_pk_cmb_CLASS_vs_SymBoltz}, but for the linear DE--DM interaction \eqref{eq:de_dm_linear}.}
    \label{fig:DE-DM-rhoDE-pk-cmb}
\end{figure}

 We consider a linear coupling between dark energy and dark matter (several examples can be found in Refs.~\cite{vanderWesthuizen:2025vcb,He:2010im,PhysRevD.89.103531,vonMarttens:2018iav,vanderWesthuizen:2023hcl,Mishra:2023ueo,Nunes:2026wfg}) defined by $Q^\nu_{\rm c}= \zeta \, \hat{\Theta} \,  T^{\alpha\beta}_{\Lambda} u^{\Lambda}_\alpha u^{\Lambda}_\beta u^\nu_{\rm c} \,/\, 3$, which translates to the following decomposition:%
 \footnote{In the code the coupling parameter always follows  the notation $\alpha{{\rm n}{\rm m}}$ where $\rm n $ and $\rm m$ are the interacting species in each case. Here, to ease the notation, we simply use $\zeta$ for this interaction which corresponds to $\zeta\equiv\alpha{\rm c}{\Lambda}$ in \SymBoltz.}
\be
\bar{Q}  = \zeta \frac{\mathcal{H}}{a} \rho_{\Lambda} , \qquad
\delta Q = \zeta \frac{\mathcal{H}}{a} \rho_{\Lambda} (\delta_{\Lambda} - \Psi)+ \zeta \frac{\rho_{\Lambda}}{a} (\tfrac13 \hat{\theta} - \Phi')  , \qquad
f        = \zeta \frac{\mathcal{H}}{k^2 a} \rho_{\Lambda} (\theta_{\rm c} - \hat{\theta}) .
\label{eq:de_dm_linear}
\ee
       Figures~\ref{fig:DE-DM-rhoDE-background} and \ref{fig:DE-DM-rhoDE-pk-cmb} again display the background evolution of the different energy densities of the interacting and non-interacting components, the matter power spectrum and the CMB power spectra for several values of the coupling parameter $\zeta$ from \SymBoltz and compare them to \CLASS.
       For the background evolution, we see that both codes agree very well up to a difference well below $0.001\%$ in the densities of the interacting components, $\rho_{\Lambda}$ and $\rho_{\rm c}$.
       Moreover, the differences for the interacting and non-interacting components are comparable.
       The resulting Hubble function depends on all the densities and agrees to a similar level as before.
       Consequently, even when the interaction affects the background evolution, the codes are in very good agreement.
       For the perturbations, the matter and CMB power spectra from both codes agree within a similar $0.1\%$--$1\%$ level as before (the spikes in the matter power spectrum errors are due to zero crossings since we are representing the absolute relative difference).
       Even for unrealistic values of the coupling, like $\zeta=1$, which have an extremely suppressed matter power spectrum and a CMB spectrum incompatible with Planck, the codes differ below or around $1\%$.
       Consequently, we have demonstrated that \SymBoltz can handle interacting models as precisely as other Boltzmann codes, and with a user-friendly way of implementing them by just writing three lines of code with the details of the interaction and with no approximation-switching schemes.

\subsubsection{Non-linear coupling}
\begin{figure}[H]
    \centering
    \vspace{-0.5cm}
    \includegraphics[width=0.9\textwidth]{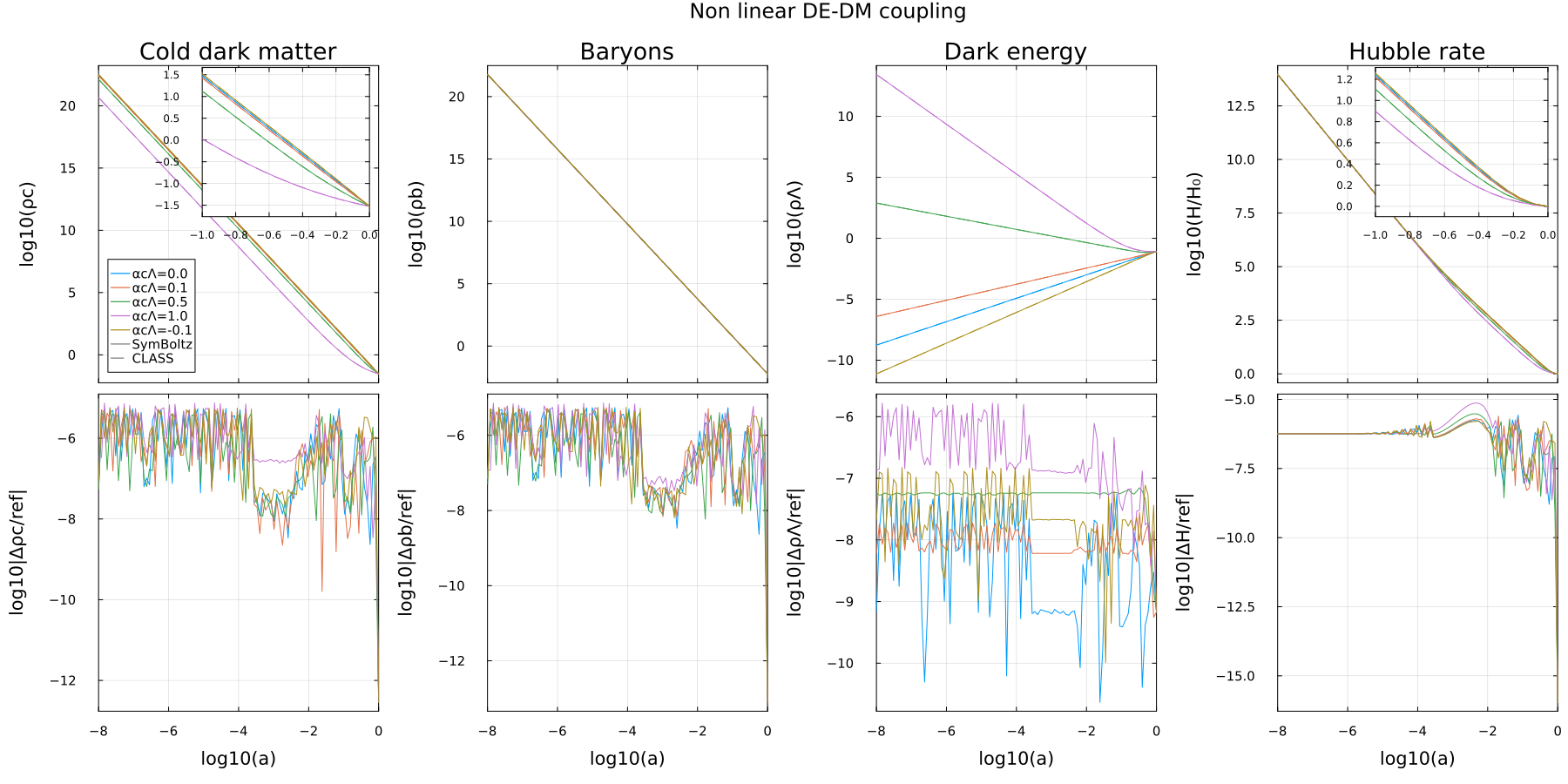}

    \caption{Background comparison between \CLASS and \SymBoltz as in Figure~\ref{fig:ref_background_CLASS_vs_SymBoltz}, but for the non-linear DE--DM interaction \eqref{eq:de_dm_nonlinear}.}
    \label{fig:DE-DM-nonlinear-background}
\end{figure}

\begin{figure}[H]
    \centering
    \vspace{-0.5cm}
    \includegraphics[height=5.9cm]{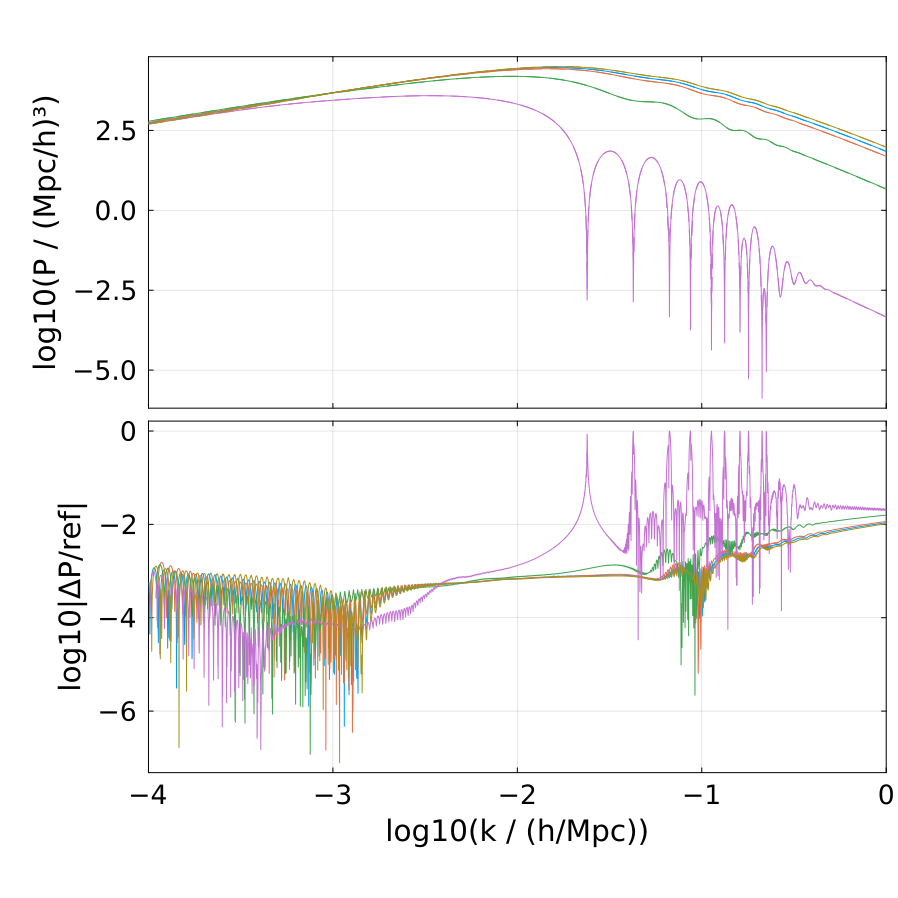}
    \includegraphics[height=5.9cm]{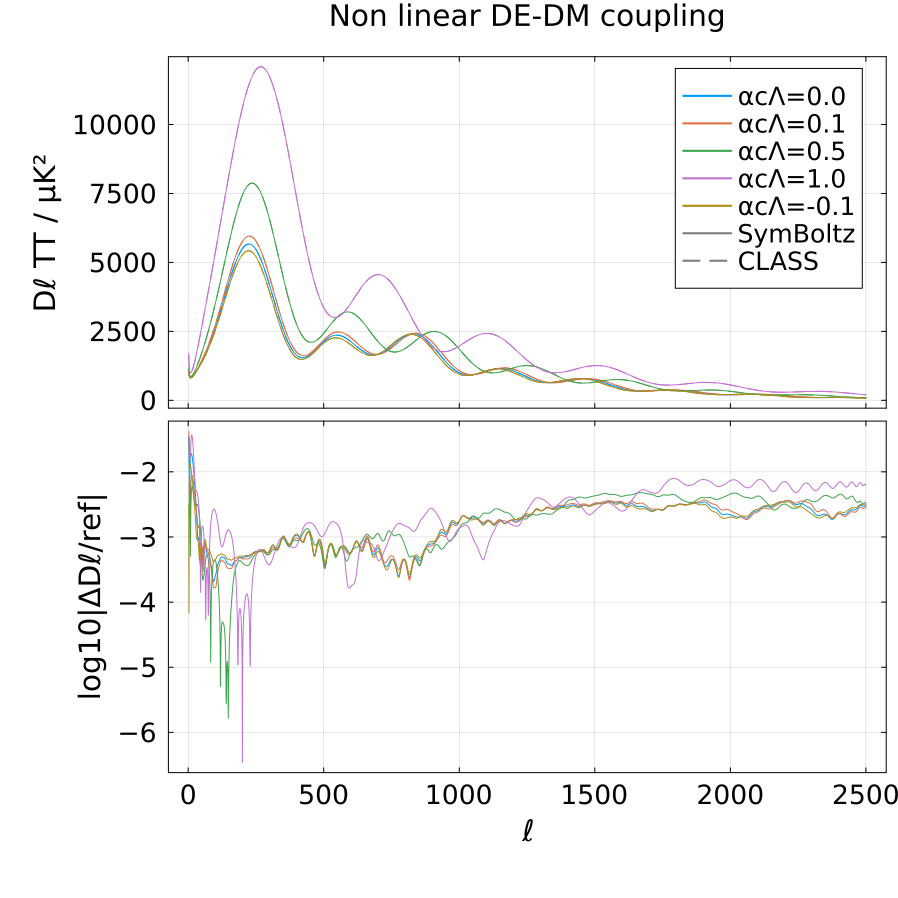}
\includegraphics[height=5.9cm]{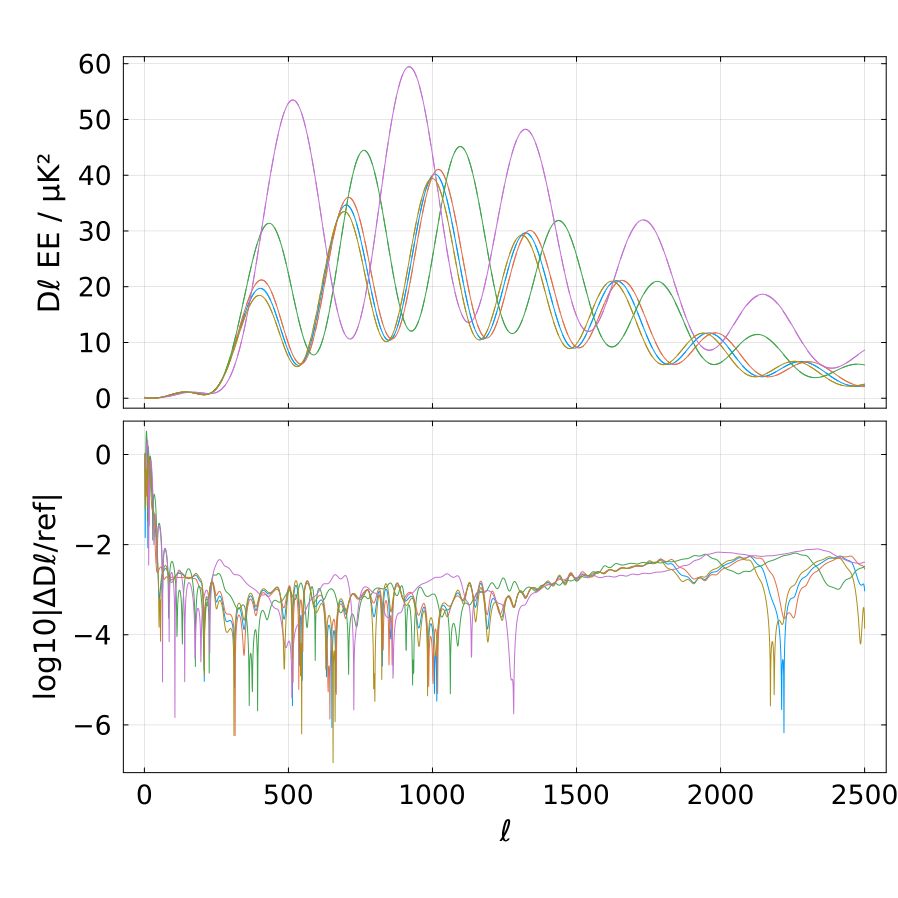}

    \caption{Power spectrum comparison between \CLASS and \SymBoltz as in Figure~\ref{fig:ref_pk_cmb_CLASS_vs_SymBoltz}, but for the non-linear DE--DM interaction \eqref{eq:de_dm_nonlinear}.}
    \label{fig:DE-DM-nonlinear-pk-cmb}
\end{figure}

 We now consider a non-linear coupling between dark energy and dark matter.
 Some examples of such models are found in Refs.~\cite{vanderWesthuizen:2025mnw,Bolotin:2013jpa,Bouhmadi-Lopez:2016dcs}.
 The interaction is defined by the four-vector
 \be
 Q^\mu_{\rm c}=\zeta \, \hat{\Theta} \, \frac{\left(T^{\alpha\beta}_{({\rm c})}u^{({\rm c})}_{\alpha}u^{({\rm c})}_{\beta}\right)\left(T^{\gamma\delta}_{({\Lambda})}u^{({\Lambda})}_{\gamma}u^{({\Lambda})}_{\delta}\right)}{T^{\alpha\beta}_{({\rm c})}u^{({\rm c})}_{\alpha}u^{({\rm c})}_{\beta}+T^{\gamma\delta}_{({\Lambda})}u^{({\Lambda})}_{\gamma}u^{({\Lambda})}_{\delta}} \, u^\mu_{\rm c} ,
 \ee
 and the decomposition gives:
\begin{subequations}
\label{eq:de_dm_nonlinear}
\begin{align}
\bar{Q}
&=
3\zeta\,\dfrac{\mathcal{H}}{a}\,
\dfrac{\rho_{\rm c}\rho_{\Lambda}}
{\rho_{\rm c}+\rho_{\Lambda}} ,
\\
\delta Q
&=
3\zeta\,\dfrac{\mathcal{H}}{a}\,
\dfrac{\rho_{\rm c}\rho_{\Lambda}}
{\rho_{\rm c}+\rho_{\Lambda}}
\left[
\dfrac{\rho_{\Lambda}\delta_{\rm c}
+\rho_{\rm c}\delta_{\Lambda}}
{\rho_{\rm c}+\rho_{\Lambda}}
-\Psi
\right]
+\zeta\,\dfrac{1}{a}\,
\dfrac{\rho_{\rm c}\rho_{\Lambda}}
{\rho_{\rm c}+\rho_{\Lambda}}
\left(
\hat{\theta}-3\Phi'
\right) ,
\\
f
&=
3\zeta\,\dfrac{\mathcal{H}}{k^2a}\,
\dfrac{\rho_{\rm c}\rho_{\Lambda}}
{\rho_{\rm c}+\rho_{\Lambda}}
\left(
\theta_{\rm c}-\hat{\theta}
\right) .
\end{align}
\end{subequations}
Figures~\ref{fig:DE-DM-nonlinear-background} and~\ref{fig:DE-DM-nonlinear-pk-cmb} again compare the background and matter and CMB power spectra from \SymBoltz and \CLASS for several values of the coupling $\zeta$.
Although the modifications produced by this interaction are phenomenologically similar to those obtained in the linear case, studying this model provides a non-trivial additional test.
The energy-transfer rate now depends non-linearly on the background densities, and both codes must consistently evaluate this non-linear dependence and its corresponding first-order perturbation throughout the cosmological evolution.
In the background, we again see that the densities and Hubble function agree well below $0.001\%$.
The power spectra also agree very well within $0.1\%$--$1\%$ for almost all scales and multipoles in a similar way as before.
This demonstrates that \SymBoltz handles non-linear interacting models as precisely as other Boltzmann codes.

This validation of several interacting models establishes \SymBoltz as a precise and powerful framework for studying interacting cosmologies, while also making it simple to modify and extend.
Any new interaction can be implemented by specifying only the three quantities $Q$, $\delta Q$ and $f$, each of which can be decomposed from an interaction four-vector $Q^\mu$.
In practice, incorporating a new interaction amounts to adding only a single line for each of these three quantities.
The agreement of \SymBoltz with \CLASS for standard non-interacting cosmologies, pure momentum transfer models, and both linear and non-linear energy-momentum transfer models demonstrates that the code accurately reproduces the background and perturbation evolution across a broad range of physically distinct scenarios.

\SymBoltz achieves this accuracy while solving the full stiff Einstein-Boltzmann system using implicit ODE solvers without relying on approximation switching schemes.
The absence of approximation switching schemes is particularly helpful when an interaction modifies a component that takes part in such schemes, such as the baryons.
In conventional Boltzmann codes, the interaction must also be consistently propagated through the corresponding approximation regimes and switching conditions.
Most importantly, this capability not only works for the standard model, but is preserved when the cosmological model is modified.
A new interaction is introduced directly through the three quantities $Q$, $\delta Q$ and $f$, which are incorporated into the same general system of equations.
Thus, \SymBoltz combines the precision required for cosmological predictions with a flexible and easy-to-modify structure that makes the exploration of general interacting models direct, systematic, and considerably simpler than modifying a conventional Boltzmann code.

\section{Multiple interactions at the same time}
\label{sec:multiple_interactions}
Finally, to really highlight the simplicity of \SymBoltz when dealing with modifications,
we consider a model with multiple pairwise interactions between dark energy, dark matter and baryons at the same time:%
\footnote{As previously commented, the label $\Lambda$ denotes dark energy of any form or a cosmological constant. Although an abuse of notation, we use it to keep as close as possible to the internal notation of \SymBoltz.}
\begin{subequations}
\begin{align}
Q_{\rm bc}^{\nu}
&=
\alpha_{\rm bc}\frac{\hat{\Theta}}{3}
\left[
T^{\alpha\beta}_{\rm (b)}u^{\rm (b)}_{\alpha}u^{\rm (b)}_{\beta}
+
T^{\alpha\beta}_{\rm (c)}u^{\rm (c)}_{\alpha}u^{\rm (c)}_{\beta}
\right]
{u}^{\nu}_{\rm c} ,
\\
Q_{{\rm b}\Lambda}^{\nu}
&=
\alpha_{{\rm b}\Lambda}\frac{\hat{\Theta}}{3}
\left[
T^{\alpha\beta}_{\rm (b)}u^{\rm (b)}_{\alpha}u^{(b)}_{\beta}
+
T^{\alpha\beta}_{(\Lambda)}u^{(\Lambda)}_{\alpha}u^{(\Lambda)}_{\beta}
\right]
{u}^{\nu}_{\rm b} ,
\\
Q_{{\rm c}\Lambda}^{\nu}
&=
\alpha_{{\rm c}\Lambda}\frac{\hat{\Theta}}{3}
\left[
T^{\alpha\beta}_{\rm (c)}u^{\rm (c)}_{\alpha}u^{\r (c)}_{\beta}
+
T^{\alpha\beta}_{(\Lambda)}u^{(\Lambda)}_{\alpha}u^{(\Lambda)}_{\beta}
\right]
{u}^{\nu}_{\rm c} .
\end{align}
\end{subequations}
Following the general decomposition in Section~\ref{subsec:cov_couplings} for each of the interactions now gives:
\begin{subequations}\label{eq:array_final_boss}
\begin{equation*}
\renewcommand{\arraystretch}{2.45}
\begin{array}{|>{\displaystyle}c|>{\displaystyle}c|>{\displaystyle}c|@{}>{\quad}r@{}}
\cline{1-3}
\bar Q_{\rm(n)(m)}
&
\delta Q_{\rm(n)(m)}
&
f_{\rm(n)(m)}
&
\\ \cline{1-3}
\rule[-3em]{0pt}{6.6em}
\bar Q_{\rm bc}
=
\alpha_{\rm bc}\frac{\mathcal H}{a}
\left(\bar\rho_{\rm b}+\bar\rho_{\rm c}\right)
&
\begin{aligned}
\delta Q_{\rm bc}
&=
\alpha_{\rm bc}\frac{\mathcal H}{a}
\left[
\bar\rho_{\rm b}\delta_{\rm b}+\bar\rho_{\rm c}\delta_{\rm c}
-\left(\bar\rho_{\rm b}+\bar\rho_{\rm c}\right)\Psi
\right]
\\
&\quad+
\alpha_{\rm bc}\frac{\bar\rho_{\rm b}+\bar\rho_{\rm c}}{a}
\left(\tfrac13 \hat\theta - \Phi'\right)
\end{aligned}
&
f_{\rm bc}
=
\alpha_{\rm bc}\frac{\mathcal H}{a k^2}
\left(\bar\rho_{\rm b}+\bar\rho_{\rm c}\right)
\left(\theta_{\rm c}-\hat\theta\right)
&
\rowtag
\\ \cline{1-3}
\rule[-3em]{0pt}{6.6em}
\bar Q_{{\rm b}\Lambda}
=
\alpha_{{\rm b}\Lambda}\frac{\mathcal H}{a}
\left(\bar\rho_{\rm b}+\bar\rho_\Lambda\right)
&
\begin{aligned}
\delta Q_{{\rm b}\Lambda}
&=
\alpha_{{\rm b}\Lambda}\frac{\mathcal H}{a}
\left[
\bar\rho_{\rm b}\delta_{\rm b}+\bar\rho_\Lambda\delta_\Lambda
-\left(\bar\rho_{\rm b}+\bar\rho_\Lambda\right)\Psi
\right]
\\
&\quad+
\alpha_{{\rm b}\Lambda}\frac{\bar\rho_{\rm b}+\bar\rho_\Lambda}{a}
\left(\tfrac13 \hat\theta - \Phi'\right)
\end{aligned}
&
f_{{\rm b}\Lambda}
=
\alpha_{{\rm b}\Lambda}\frac{\mathcal H}{a k^2}
\left(\bar\rho_{\rm b}+\bar\rho_\Lambda\right)
\left(\theta_{\rm b}-\hat\theta\right)
&
\rowtag
\\ \cline{1-3}
\rule[-3em]{0pt}{6.6em}
\bar Q_{{\rm c}\Lambda}
=
\alpha_{{\rm c}\Lambda}\frac{\mathcal H}{a}
\left(\bar\rho_{\rm c}+\bar\rho_\Lambda\right)
&
\begin{aligned}
\delta Q_{{\rm c}\Lambda}
&=
\alpha_{{\rm c}\Lambda}\frac{\mathcal H}{a}
\left[
\bar\rho_{\rm c}\delta_{\rm c}+\bar\rho_\Lambda\delta_\Lambda
-\left(\bar\rho_{\rm c}+\bar\rho_\Lambda\right)\Psi
\right]
\\
&\quad+
\alpha_{{\rm c}\Lambda}\frac{\bar\rho_{\rm c}+\bar\rho_\Lambda}{a}
\left(\tfrac13 \hat\theta - \Phi'\right)
\end{aligned}
&
f_{{\rm c}\Lambda}
=
\alpha_{{\rm c}\Lambda}\frac{\mathcal H}{a k^2}
\left(\bar\rho_{\rm c}+\bar\rho_\Lambda\right)
\left(\theta_{\rm c}-\hat\theta\right)
&
\rowtag
\\ \cline{1-3}
\end{array}
\end{equation*}
\end{subequations}
We continue to use the sign convention \eqref{eq:interaction-sign-convention} to guarantee the conservation of the total energy-momentum \eqref{eq:Qsumzero}.
This gives the total $\Qb$, $\delta Q$ and $f$ ``influxes'' to each component:
\begin{subequations}
\begin{align}
\bar Q_{\rm b}&=-\bar Q_{\rm bc}+\bar Q_{{\rm b}\Lambda},
&
\delta Q_{\rm b}&=-\delta Q_{\rm bc}+\delta Q_{{\rm b}\Lambda},
&
f_{\rm b}&=-f_{\rm bc}+f_{{\rm b}\Lambda},
\\
\bar Q_{\rm c}&=+\bar Q_{\rm bc}+\bar Q_{{\rm c}\Lambda},
&
\delta Q_{\rm c}&=+\delta Q_{\rm bc}+\delta Q_{{\rm c}\Lambda},
&
f_{\rm c}&=+f_{\rm bc}+f_{{\rm c}\Lambda},
\\
\bar Q_\Lambda&=-\bar Q_{{\rm b}\Lambda}-\bar Q_{{\rm c}\Lambda},
&
\delta Q_\Lambda&=-\delta Q_{{\rm b}\Lambda}-\delta Q_{{\rm c}\Lambda},
&
f_\Lambda&=-f_{{\rm b}\Lambda}-f_{{\rm c}\Lambda}.
\end{align}
\end{subequations}
This guarantees cancellation of the energy and momentum transfers when all components are added together,
so the total stress-energy tensor is covariantly conserved for the interacting components both in the background and perturbations.
That is,
$
\nabla_\nu\big(
T^{\mu\nu}_{\rm (b)}
+
T^{\mu\nu}_{\rm (c)}
+
T^{\mu\nu}_{(\Lambda)}
\big)
=
Q^\mu_{({\rm  b})}
+
Q^\mu_{({\rm c})}
+
Q^\mu_{({\Lambda})}
= 0 .
$

This complicated system is implemented in \SymBoltz in the very same simple way as all the previous test cases.
By just writing the $3\times3$ expressions \eqref{eq:array_final_boss} for the $\Qb_{\rm(n)(m)}$, $\delta Q_{\rm(n)(m)}$ and $f_{\rm(n)(m)}$ into the ``factory function'' \verb|interacting_model()| as explained in Section~\ref{sec:implementation}, \SymBoltz automatically creates the correct background and perturbation equations with no other necessary modifications from the user.
Then the user only needs to specify the desired values of the coupling parameters $\alpha_{\rm nm}$ and solve the resulting background and perturbation equations.
Taking the interactions \eqref{eq:array_final_boss} as an example model and a proof of concept for the utility of \SymBoltz, the user only has to write the following code:
\newpage
\begin{snugshade}
\begin{Verbatim}[commandchars=\\\{\},fontsize=\small,breaklines,xleftmargin=5pt,xrightmargin=5pt]
\PY{n}{M}\PY{+w}{ }\PY{o}{=}\PY{+w}{ }\PY{n}{interacting\PYZus{}model}\PY{p}{(}\PY{p}{)}
\PY{n+nd}{@unpack}\PY{+w}{ }\PY{n}{a}\PY{p}{,}\PY{+w}{ }\PY{n}{ℋ}\PY{p}{,}\PY{+w}{ }\PY{n}{ρb}\PY{p}{,}\PY{+w}{ }\PY{n}{ρc}\PY{p}{,}\PY{+w}{ }\PY{n}{ρΛ}\PY{p}{,}\PY{+w}{ }\PY{n}{δb}\PY{p}{,}\PY{+w}{ }\PY{n}{δc}\PY{p}{,}\PY{+w}{ }\PY{n}{δΛ}\PY{p}{,}\PY{+w}{ }\PY{n}{θ}\PY{p}{,}\PY{+w}{ }\PY{n}{θb}\PY{p}{,}\PY{+w}{ }\PY{n}{θc}\PY{p}{,}\PY{+w}{ }\PY{n}{θΛ}\PY{p}{,}\PY{+w}{ }\PY{n}{Ψ}\PY{p}{,}\PY{+w}{ }\PY{n}{Φ}\PY{p}{,}\PY{+w}{ }\PY{n}{k}\PY{p}{,}\PY{+w}{ }\PY{n}{αbc}\PY{p}{,}\PY{+w}{ }\PY{n}{αbΛ}\PY{p}{,}\PY{+w}{ }\PY{n}{αcΛ}\PY{+w}{ }\PY{o}{=}\PY{+w}{ }\PY{n}{M}

\PY{c}{\PYZsh{} Background energy transfers}
\PY{n}{Qbc}\PY{+w}{ }\PY{o}{=}\PY{+w}{ }\PY{n}{αbc}\PY{+w}{ }\PY{o}{*}\PY{+w}{ }\PY{n}{ℋ}\PY{+w}{ }\PY{o}{/}\PY{+w}{ }\PY{n}{a}\PY{+w}{ }\PY{o}{*}\PY{+w}{ }\PY{p}{(}\PY{n}{ρb}\PY{+w}{ }\PY{o}{+}\PY{+w}{ }\PY{n}{ρc}\PY{p}{)}
\PY{n}{QbΛ}\PY{+w}{ }\PY{o}{=}\PY{+w}{ }\PY{n}{αbΛ}\PY{+w}{ }\PY{o}{*}\PY{+w}{ }\PY{n}{ℋ}\PY{+w}{ }\PY{o}{/}\PY{+w}{ }\PY{n}{a}\PY{+w}{ }\PY{o}{*}\PY{+w}{ }\PY{p}{(}\PY{n}{ρb}\PY{+w}{ }\PY{o}{+}\PY{+w}{ }\PY{n}{ρΛ}\PY{p}{)}
\PY{n}{QcΛ}\PY{+w}{ }\PY{o}{=}\PY{+w}{ }\PY{n}{αcΛ}\PY{+w}{ }\PY{o}{*}\PY{+w}{ }\PY{n}{ℋ}\PY{+w}{ }\PY{o}{/}\PY{+w}{ }\PY{n}{a}\PY{+w}{ }\PY{o}{*}\PY{+w}{ }\PY{p}{(}\PY{n}{ρc}\PY{+w}{ }\PY{o}{+}\PY{+w}{ }\PY{n}{ρΛ}\PY{p}{)}

\PY{c}{\PYZsh{} Perturbations of the energy transfers}
\PY{n}{δQbc}\PY{+w}{ }\PY{o}{=}\PY{+w}{ }\PY{n}{αbc}\PY{+w}{ }\PY{o}{*}\PY{+w}{ }\PY{n}{ℋ}\PY{+w}{ }\PY{o}{/}\PY{+w}{ }\PY{n}{a}\PY{+w}{ }\PY{o}{*}\PY{+w}{ }\PY{p}{(}\PY{n}{ρb}\PY{+w}{ }\PY{o}{*}\PY{+w}{ }\PY{n}{δb}\PY{+w}{ }\PY{o}{+}\PY{+w}{ }\PY{n}{ρc}\PY{+w}{ }\PY{o}{*}\PY{+w}{ }\PY{n}{δc}\PY{+w}{ }\PY{o}{\PYZhy{}}\PY{+w}{ }\PY{p}{(}\PY{n}{ρb}\PY{+w}{ }\PY{o}{+}\PY{+w}{ }\PY{n}{ρc}\PY{p}{)}\PY{+w}{ }\PY{o}{*}\PY{+w}{ }\PY{n}{Ψ}\PY{p}{)}\PY{+w}{ }\PY{o}{+}\PY{+w}{ }\PY{n}{αbc}\PY{+w}{ }\PY{o}{*}\PY{+w}{ }\PY{p}{(}\PY{n}{ρb}\PY{+w}{ }\PY{o}{+}\PY{+w}{ }\PY{n}{ρc}\PY{p}{)}\PY{+w}{ }\PY{o}{/}\PY{+w}{ }\PY{n}{a}\PY{+w}{ }\PY{o}{*}\PY{+w}{ }\PY{p}{(}\PY{n}{θ}\PY{+w}{ }\PY{o}{/}\PY{+w}{ }\PY{l+m+mi}{3}\PY{+w}{ }\PY{o}{\PYZhy{}}\PY{+w}{ }\PY{n}{Dτ}\PY{p}{(}\PY{n}{Φ}\PY{p}{)}\PY{p}{)}
\PY{n}{δQbΛ}\PY{+w}{ }\PY{o}{=}\PY{+w}{ }\PY{n}{αbΛ}\PY{+w}{ }\PY{o}{*}\PY{+w}{ }\PY{n}{ℋ}\PY{+w}{ }\PY{o}{/}\PY{+w}{ }\PY{n}{a}\PY{+w}{ }\PY{o}{*}\PY{+w}{ }\PY{p}{(}\PY{n}{ρb}\PY{+w}{ }\PY{o}{*}\PY{+w}{ }\PY{n}{δb}\PY{+w}{ }\PY{o}{+}\PY{+w}{ }\PY{n}{ρΛ}\PY{+w}{ }\PY{o}{*}\PY{+w}{ }\PY{n}{δΛ}\PY{+w}{ }\PY{o}{\PYZhy{}}\PY{+w}{ }\PY{p}{(}\PY{n}{ρb}\PY{+w}{ }\PY{o}{+}\PY{+w}{ }\PY{n}{ρΛ}\PY{p}{)}\PY{+w}{ }\PY{o}{*}\PY{+w}{ }\PY{n}{Ψ}\PY{p}{)}\PY{+w}{ }\PY{o}{+}\PY{+w}{ }\PY{n}{αbΛ}\PY{+w}{ }\PY{o}{*}\PY{+w}{ }\PY{p}{(}\PY{n}{ρb}\PY{+w}{ }\PY{o}{+}\PY{+w}{ }\PY{n}{ρΛ}\PY{p}{)}\PY{+w}{ }\PY{o}{/}\PY{+w}{ }\PY{n}{a}\PY{+w}{ }\PY{o}{*}\PY{+w}{ }\PY{p}{(}\PY{n}{θ}\PY{+w}{ }\PY{o}{/}\PY{+w}{ }\PY{l+m+mi}{3}\PY{+w}{ }\PY{o}{\PYZhy{}}\PY{+w}{ }\PY{n}{Dτ}\PY{p}{(}\PY{n}{Φ}\PY{p}{)}\PY{p}{)}
\PY{n}{δQcΛ}\PY{+w}{ }\PY{o}{=}\PY{+w}{ }\PY{n}{αcΛ}\PY{+w}{ }\PY{o}{*}\PY{+w}{ }\PY{n}{ℋ}\PY{+w}{ }\PY{o}{/}\PY{+w}{ }\PY{n}{a}\PY{+w}{ }\PY{o}{*}\PY{+w}{ }\PY{p}{(}\PY{n}{ρc}\PY{+w}{ }\PY{o}{*}\PY{+w}{ }\PY{n}{δc}\PY{+w}{ }\PY{o}{+}\PY{+w}{ }\PY{n}{ρΛ}\PY{+w}{ }\PY{o}{*}\PY{+w}{ }\PY{n}{δΛ}\PY{+w}{ }\PY{o}{\PYZhy{}}\PY{+w}{ }\PY{p}{(}\PY{n}{ρc}\PY{+w}{ }\PY{o}{+}\PY{+w}{ }\PY{n}{ρΛ}\PY{p}{)}\PY{+w}{ }\PY{o}{*}\PY{+w}{ }\PY{n}{Ψ}\PY{p}{)}\PY{+w}{ }\PY{o}{+}\PY{+w}{ }\PY{n}{αcΛ}\PY{+w}{ }\PY{o}{*}\PY{+w}{ }\PY{p}{(}\PY{n}{ρc}\PY{+w}{ }\PY{o}{+}\PY{+w}{ }\PY{n}{ρΛ}\PY{p}{)}\PY{+w}{ }\PY{o}{/}\PY{+w}{ }\PY{n}{a}\PY{+w}{ }\PY{o}{*}\PY{+w}{ }\PY{p}{(}\PY{n}{θ}\PY{+w}{ }\PY{o}{/}\PY{+w}{ }\PY{l+m+mi}{3}\PY{+w}{ }\PY{o}{\PYZhy{}}\PY{+w}{ }\PY{n}{Dτ}\PY{p}{(}\PY{n}{Φ}\PY{p}{)}\PY{p}{)}

\PY{c}{\PYZsh{} Scalar momentum\PYZhy{}transfer potentials}
\PY{n}{fQbc}\PY{+w}{ }\PY{o}{=}\PY{+w}{ }\PY{n}{αbc}\PY{+w}{ }\PY{o}{*}\PY{+w}{ }\PY{n}{ℋ}\PY{+w}{ }\PY{o}{/}\PY{+w}{ }\PY{p}{(}\PY{n}{a}\PY{+w}{ }\PY{o}{*}\PY{+w}{ }\PY{n}{k}\PY{o}{\PYZca{}}\PY{l+m+mi}{2}\PY{p}{)}\PY{+w}{ }\PY{o}{*}\PY{+w}{ }\PY{p}{(}\PY{n}{ρb}\PY{+w}{ }\PY{o}{+}\PY{+w}{ }\PY{n}{ρc}\PY{p}{)}\PY{+w}{ }\PY{o}{*}\PY{+w}{ }\PY{p}{(}\PY{n}{θc}\PY{+w}{ }\PY{o}{\PYZhy{}}\PY{+w}{ }\PY{n}{θ}\PY{p}{)}
\PY{n}{fQbΛ}\PY{+w}{ }\PY{o}{=}\PY{+w}{ }\PY{n}{αbΛ}\PY{+w}{ }\PY{o}{*}\PY{+w}{ }\PY{n}{ℋ}\PY{+w}{ }\PY{o}{/}\PY{+w}{ }\PY{p}{(}\PY{n}{a}\PY{+w}{ }\PY{o}{*}\PY{+w}{ }\PY{n}{k}\PY{o}{\PYZca{}}\PY{l+m+mi}{2}\PY{p}{)}\PY{+w}{ }\PY{o}{*}\PY{+w}{ }\PY{p}{(}\PY{n}{ρb}\PY{+w}{ }\PY{o}{+}\PY{+w}{ }\PY{n}{ρΛ}\PY{p}{)}\PY{+w}{ }\PY{o}{*}\PY{+w}{ }\PY{p}{(}\PY{n}{θb}\PY{+w}{ }\PY{o}{\PYZhy{}}\PY{+w}{ }\PY{n}{θ}\PY{p}{)}
\PY{n}{fQcΛ}\PY{+w}{ }\PY{o}{=}\PY{+w}{ }\PY{n}{αcΛ}\PY{+w}{ }\PY{o}{*}\PY{+w}{ }\PY{n}{ℋ}\PY{+w}{ }\PY{o}{/}\PY{+w}{ }\PY{p}{(}\PY{n}{a}\PY{+w}{ }\PY{o}{*}\PY{+w}{ }\PY{n}{k}\PY{o}{\PYZca{}}\PY{l+m+mi}{2}\PY{p}{)}\PY{+w}{ }\PY{o}{*}\PY{+w}{ }\PY{p}{(}\PY{n}{ρc}\PY{+w}{ }\PY{o}{+}\PY{+w}{ }\PY{n}{ρΛ}\PY{p}{)}\PY{+w}{ }\PY{o}{*}\PY{+w}{ }\PY{p}{(}\PY{n}{θc}\PY{+w}{ }\PY{o}{\PYZhy{}}\PY{+w}{ }\PY{n}{θ}\PY{p}{)}

\PY{n}{M1}\PY{+w}{ }\PY{o}{=}\PY{+w}{ }\PY{n}{interacting\PYZus{}model}\PY{p}{(}\PY{p}{;}\PY{+w}{ }\PY{n}{Qbc}\PY{p}{,}\PY{+w}{ }\PY{n}{QbΛ}\PY{p}{,}\PY{+w}{ }\PY{n}{QcΛ}\PY{p}{,}\PY{+w}{ }\PY{n}{δQbc}\PY{p}{,}\PY{+w}{ }\PY{n}{δQbΛ}\PY{p}{,}\PY{+w}{ }\PY{n}{δQcΛ}\PY{p}{,}\PY{+w}{ }\PY{n}{fQbc}\PY{p}{,}\PY{+w}{ }\PY{n}{fQbΛ}\PY{p}{,}\PY{+w}{ }\PY{n}{fQcΛ}\PY{p}{)}
\end{Verbatim}
\end{snugshade}
\begin{figure}[H]
    \centering
    \includegraphics[width=0.9\textwidth]{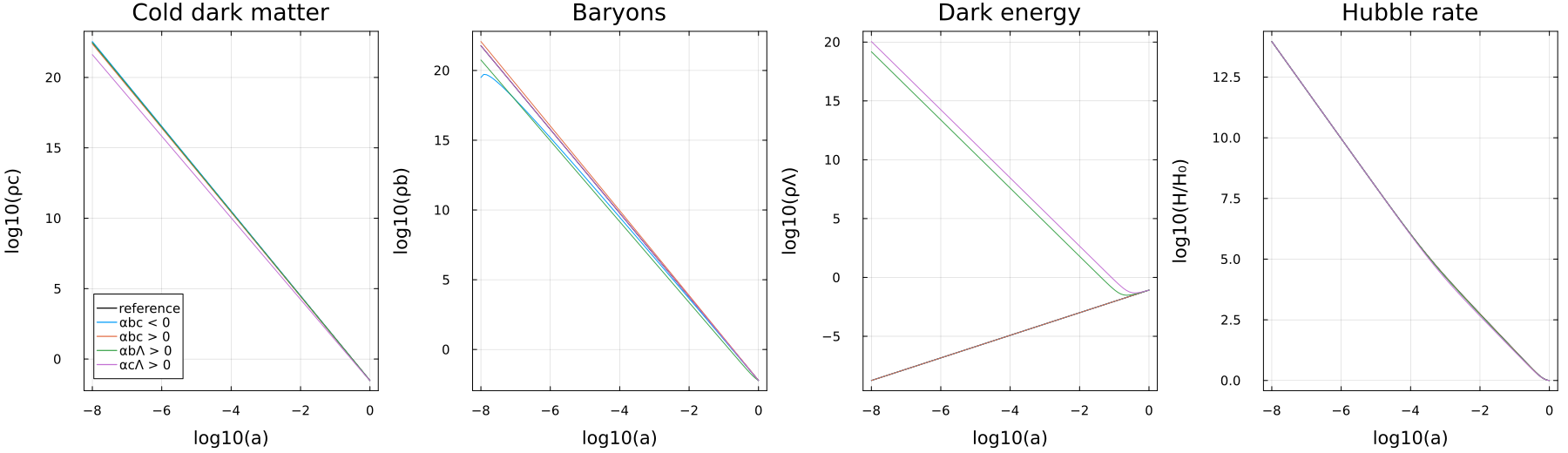}
    \includegraphics[width=0.9\textwidth]{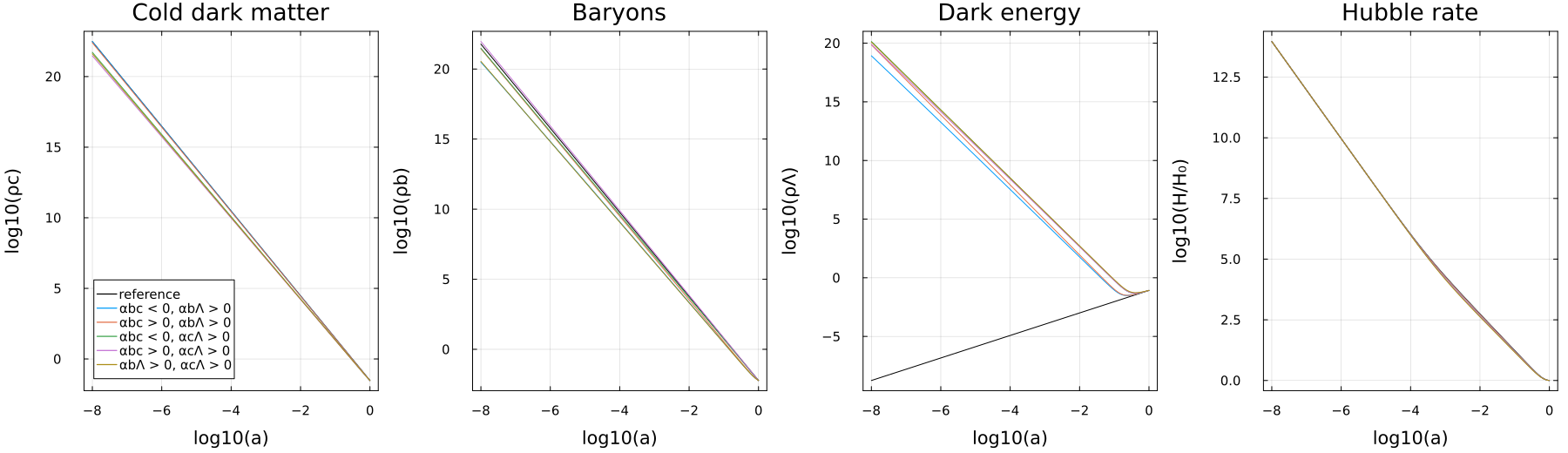}
    \includegraphics[width=0.9\textwidth]{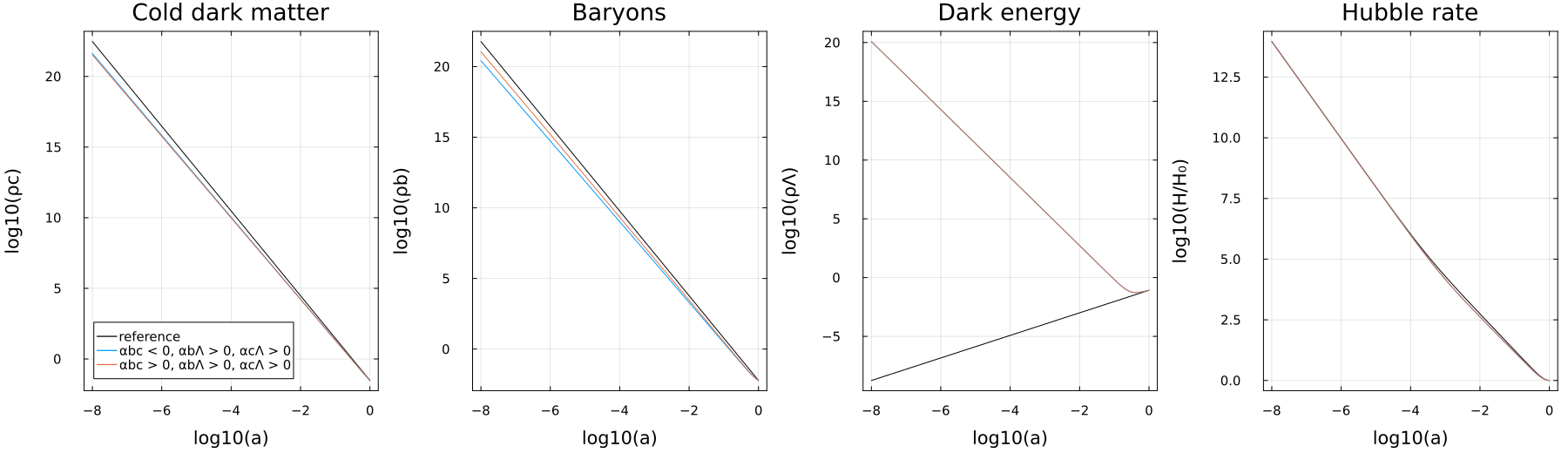}
    \caption{Background evolution for the multi-interacting baryon--dark matter--dark energy model \eqref{eq:array_final_boss}. From top to bottom: one individual interaction, two simultaneous interactions, and the three interactions active at the same time.}
    \label{fig:final_boss_background}
\end{figure}

\begin{figure}[H]
    \centering
    \includegraphics[width=0.325\textwidth]{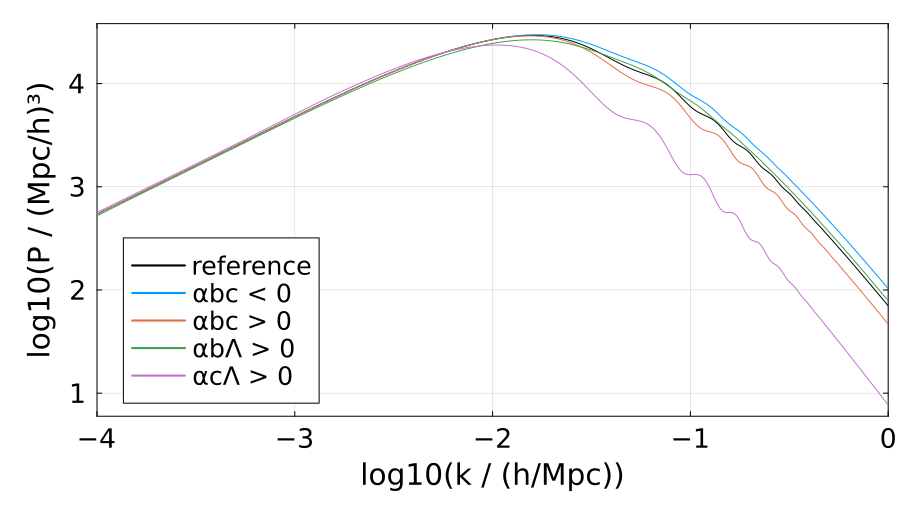}
    \includegraphics[width=0.325\textwidth]{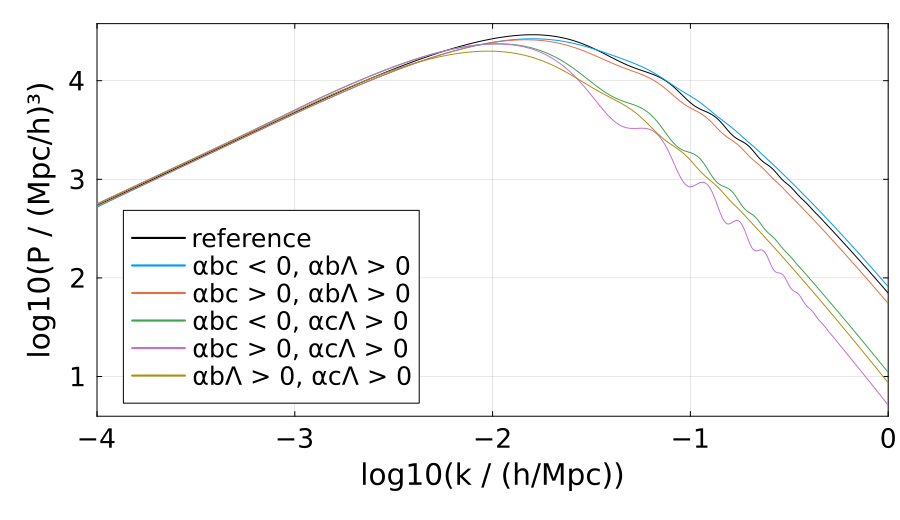}    
    \includegraphics[width=0.325\textwidth]{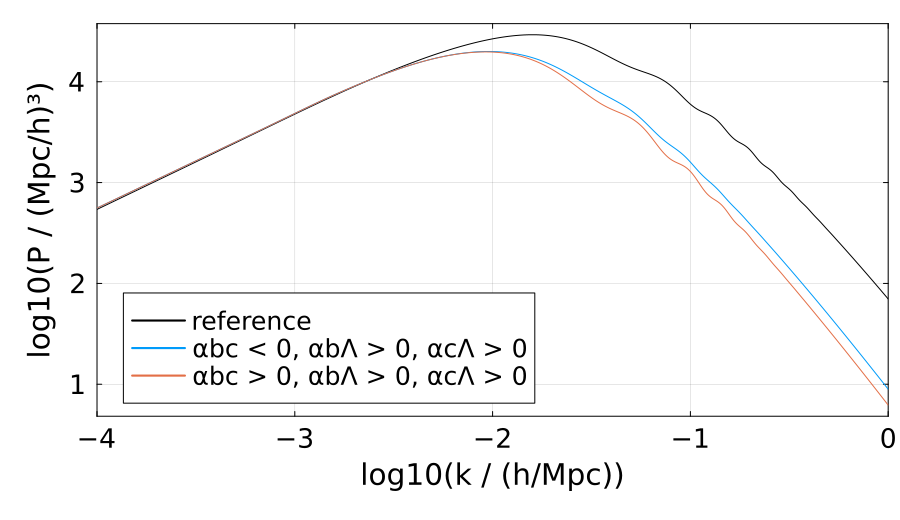}
    \caption{Matter power spectrum for the multi-interacting baryon--dark matter--dark energy model \eqref{eq:array_final_boss}. From left to right: one individual interaction, two simultaneous interactions, and all three interactions active at the same time.}
    \label{fig:final_boss_pk}
\end{figure}

\begin{figure}[H]
    \centering
    \includegraphics[width=0.9\textwidth]{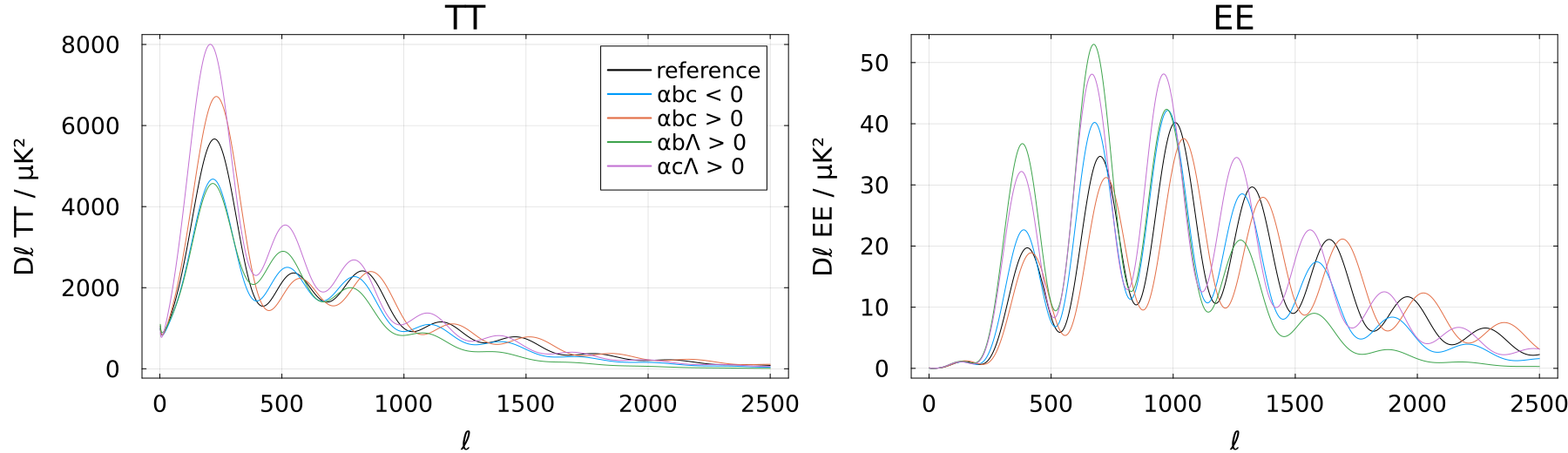}    
    \includegraphics[width=0.9\textwidth]{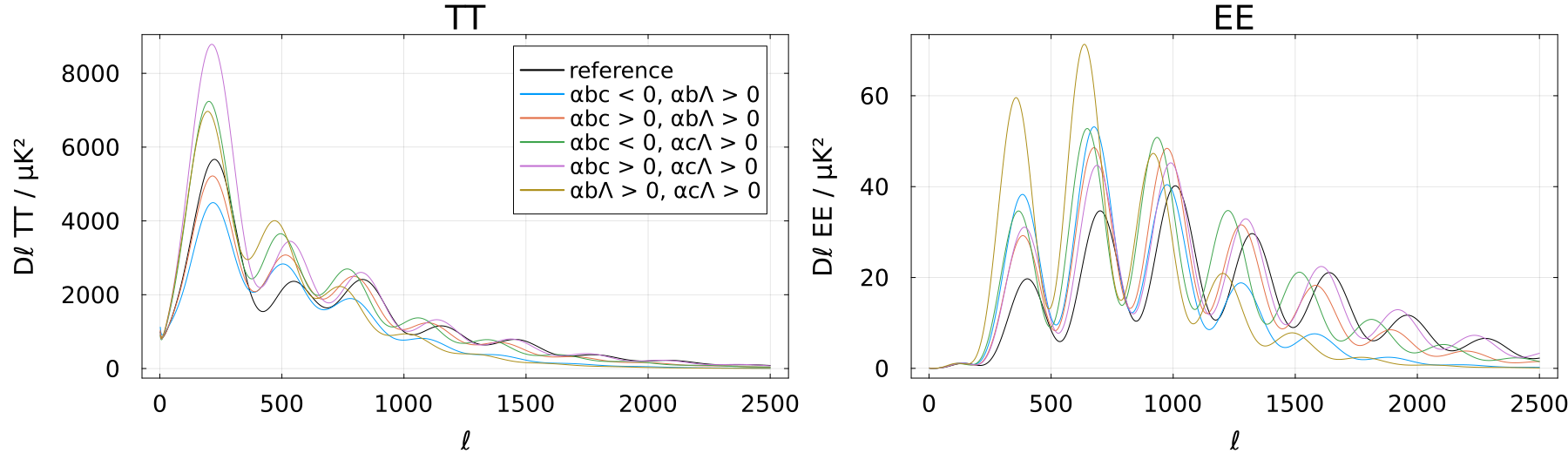}    
    \includegraphics[width=0.9\textwidth]{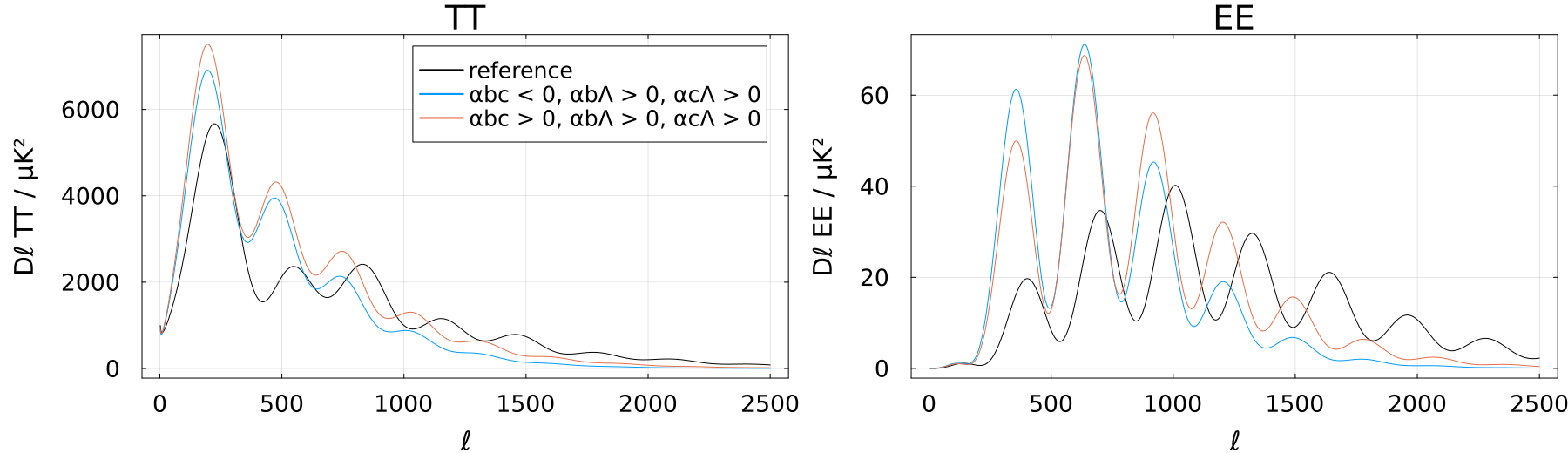}
    \caption{CMB TT and EE angular power spectra for the multi-interacting baryon--dark matter--dark energy model \eqref{eq:array_final_boss}. From top to bottom: one individual interaction, two simultaneous interactions, and all three interactions active at the same time.}
    \label{fig:final_boss_cmb}
\end{figure}

Once the three pairwise interactions have been easily implemented in \SymBoltz, we now consider three increasingly complex cases with one, two and three interactions active at a time.
For that we only need to set the values of the coupling parameters $\alpha$\texttt{bc}, $\alpha$\texttt{b}$\Lambda$ and $\alpha$\texttt{c}$\Lambda$ in the code.
Figure~\ref{fig:final_boss_background} displays the background evolution with densities and the Hubble rate for each case.
Figures~\ref{fig:final_boss_pk} and~\ref{fig:final_boss_cmb} display the matter and CMB power spectra, respectively.
Here we have no corresponding implementation in \CLASS to compare with,
which would take us more effort and care to build, particularly for the case of the interactions involving baryons where we would need to propagate the new coupling terms into the TCA schemes. All that complexity is absent here. 
This example shows how \SymBoltz makes it relatively simple to implement such complex models.

\section{Conclusion}
\label{sec:conclusions}
The main goal of this work is to show how the Einstein--Boltzmann code \SymBoltz
makes it particularly easy to implement interactions between baryons, dark matter and dark energy in a general, direct and simple way.
Starting from \textnormal{SymBoltz}' compact symbolic definition of the full standard \LCDM{} model,
we created a simple $\texttt{interacting\_model(}Q_{\rm(n)(m)}, \delta Q_{\rm(n)(m)},f_{\rm(n)(m)}\texttt{)}$ wrapper function around it that allows specifying arbitrary expressions for the flow of energy and momentum between any pair(s) of fluids $\{\rm(n)(m)\}$.
This makes it trivial to prototype arbitrary interactions between any fluids,
and in a way where multiple interactions are not much more complex than only one.
Our full model definition is available in the code documentation and is easy to modify.%
\footnote{\url{https://hersle.github.io/SymBoltz.jl/stable/interacting/}}

Of course, the interaction should be physically well-defined and stable.
Here we showed how to derive interactions consistently from a covariant formulation
and in a way that conserves total energy-momentum by construction.
Once this is done, the implementation in \SymBoltz is straightforward
and only involves typing the symbolic expressions into the code,
with no additional boilerplate code.

The symbolic-numeric interface and approximation-free nature of \SymBoltz make it possible to write all variables, parameters and equations of the cosmological model in a convenient, uniform and compact modeling language.
Support for Unicode characters makes the code equations readable and correspond one-to-one with the mathematical notation.
All equations can be specified in a single place \emph{as if} they are part of one system,
but \SymBoltz internally splits them into sequential background/perturbation stages with a mixture of forwards-backwards integration to solve them in the optimal way.
The Rodas5P implicit solver coupled with efficient generated numerical code for the ODE and its sparse and analytical Jacobian is completely invisible to users,
but this internal infrastructure makes the stiff equations stable and fast to solve without complicating the model with approximation schemes.

Concretely, here we first validated a few models with only two interacting species
by comparing them to corresponding modified versions of \CLASS from earlier works,
finding $0.1\%$--$1\%$ difference for matter and CMB power spectra at most scales.
We then constructed a proof-of-concept model with interactions between all three species at the same time.
This was completely straightforward with our general interacting symbolic model constructor,
but would take more effort and care to implement in a conventional code like \CLASS due to approximation schemes and more verbose boilerplate code for every operation in the code.

Given the flexibility of \SymBoltz, this multi-interacting model can be extended in two possible ways.
First, we can consider an interaction that not only modifies the energy-momentum conservation rule but also the thermodynamical evolution of the interacting species.
Having an effective modification on the temperature evolution equations of an interacting species can be added with a new variable, say $Q_{\rm T}$, to the triad ($Q$, $\delta Q$, $f$) for each interaction.
With the goal of having a general and univocally defined approach, one could consider a framework in which \SymBoltz takes the usual interaction triad \((\bar Q,\delta Q,f)\) together with the additional micro-physical information needed to specify the exchange of internal energy, and automatically incorporates the corresponding modifications into the temperature evolution equations through its symbolic-numeric framework.
Second, we can easily incorporate new species by adding new variables and equations in our model definition, and adding their contributions to the total energy-momentum variables that source gravity.
However, our objective here is to test the ease of modifying the code, and we leave these generalisations for future work.

To summarize, \SymBoltz provides a direct route from any arbitrarily chosen covariantly specified interaction to full, approximation-free cosmological predictions.
This makes the exploration of multi-component interacting cosmologies of any kind straightforward to code with quantitatively reliable calculations that agree very well with  established state-of-the-art Boltzmann solvers. 

\begin{acknowledgments}
The authors thank Hans Winther for useful comments and insightful discussions that helped improve this work. 
DF acknowledges support from "Convocatoria de contratación de    personal investigador doctor de la UPV/EHU (2024)" and from Project
PID2025-174372NB-I00 funded by MICIU/AEI/10.13039/501100011033.
HS acknowledges support from the Research Council of Norway under project number 325113.
\end{acknowledgments}

\bibliography{bib}

\end{document}